\documentclass[a4paper,11pt]{article}

\usepackage{jheppub}
\makeatletter
\gdef\@fpheader{}
\makeatother
\usepackage{xcolor,colortbl}
\allowdisplaybreaks

\usepackage[normalem]{ulem}
\usepackage{physics}
\usepackage{mathtools}
\usepackage{slashed}
\usepackage{subcaption}
\usepackage{dsfont}

\title{Higgs-boson production in top-quark fragmentation}

\author{Colomba Brancaccio,}
\author{Micha\l{} Czakon,}
\author{Terry Generet,}
\author{Michael Kr\"{a}mer}
\affiliation{Institut f\"ur Theoretische Teilchenphysik und Kosmologie, RWTH Aachen University,\\ D-52056 Aachen, Germany}
\emailAdd{brancaccio@physik.rwth-aachen.de}
\emailAdd{mczakon@physik.rwth-aachen.de}
\emailAdd{terry.generet@rwth-aachen.de}
\emailAdd{mkraemer@physik.rwth-aachen.de}

\abstract{We compute the fragmentation functions for the production of a Higgs boson at $\mathcal{O}(y_t^2\alpha_s)$. As part of this calculation, the relevant splitting functions are also derived at the same perturbative order. Our results can be used to compute differential cross sections with arbitrary top-quark and Higgs-boson masses from massless calculations. They can also be used to resum logarithms of the form $\ln(p_T/m)$ at large transverse momentum $p_T$ to next-to-leading-logarithmic accuracy by solving the DGLAP equations.}

\dedicated{\rm P3H-21-046, TTK-21-21}

\begin{document}
\maketitle
\flushbottom

\section{Introduction}
The associated production of a Higgs boson and a top-quark pair has been studied extensively in theory and by the experimental collaborations at the Large Hadron Collider LHC. This process is not only of interest due to the important role the Higgs boson and the top quark play in searches for models of physics beyond the Standard Model, but also because it provides a means of measuring the top-Yukawa coupling $y_t$ directly. Indeed, while currently $y_t$ is inferred from measurements by assuming the validity of the Standard Model, see e.g.\ Refs.\ \cite{Sirunyan:2020eds,Sirunyan:2018koj,Aad:2015gba}, the recent observation of the $t\overline{t}H$ final state by both the CMS \cite{Sirunyan:2018hoz} and ATLAS \cite{Aaboud:2018urx} collaborations suggests that a direct measurement of $y_t$ using this channel may soon be feasible.

Much effort has gone into improving the precision of the theoretical predictions for top-Higgs associated production: the next-to-leading order (NLO) QCD corrections have been known for a long time \cite{Beenakker:2001rj,Beenakker:2002nc,Reina:2001bc,Reina:2001sf,Dawson:2002tg,Dawson:2003zu} and have since been extended to incorporate the decay of the top quarks, including off-shell effects \cite{Denner:2015yca}. Soft-gluon resummation has been performed, first at next-to-leading-logarithmic accuracy (NLL) \cite{Kulesza:2015vda} and then also at NNLL \cite{Broggio:2016lfj,Kulesza:2020nfh,Kulesza:2017ukk}, and NNLO corrections have been computed in the soft-gluon limit \cite{Broggio:2015lya}. A NLO calculation of $t\overline{t}H$+jet is available \cite{vanDeurzen:2013xla} and NLO electroweak corrections have been calculated as well \cite{Yu:2014cka,Frixione:2014qaa,Frixione:2015zaa,Hartanto:2015uka}.

Potentially large logarithms of the form $\ln(p_T/m)$, where $m$ denotes the Higgs-boson or top-quark mass, appear at higher orders in perturbation theory. While measurements at the LHC do not currently probe the kinematic regime where such logarithms would spoil the convergence of the perturbative expansion, such regimes could become relevant for the LHC in the future and will be important at  future colliders with higher partonic centre-of-mass energies. The resummation of mass logarithms can be performed using the perturbative fragmentation function formalism \cite{Mele:1990cw}. The LO and NLO top-Higgs fragmentation functions have been presented in Ref.~\cite{Dawson:1997im} in the limit $m_H^2 \ll m_t^2$ and based on a soft-gluon approximation. A complete LO calculation of the top-Higgs fragmentation function based on QCD factorisation and including the full mass dependence has been performed in Ref.~\cite{Braaten:2015ppa}. 
 
A further important benefit of the fragmentation function formalism is that in order to obtain the fully massive cross section in the high-$p_T$ regime, the partonic cross sections need only be known in the massless limit $m_t=m_H=0$, considerably simplifying the calculation. While this is only of academic value at NLO, as the NLO corrections are known for general values of the masses, it may be of significant value when attempting to compute the NNLO corrections to associated $t\bar{t}H$ production. As such, the calculation within the fragmentation function formalism at NLO may be a stepping stone towards the full NNLO corrections at high $p_T$. Of course, in order to perform such computations one needs to know the relevant fragmentation functions (and time-like splitting functions) at the same order in perturbation theory.

In this paper we present the first complete calculation of the $\mathcal{O}(y_t^2\alpha_s)$ corrections to the perturbative fragmentation functions and to the splitting functions relevant for associated top-Higgs production, namely those describing the final-state transitions $t\to H$ and $g\to H$. The paper is organised as follows. The fragmentation function formalism is briefly presented in Section~\ref{sec:PFF}. In Section~\ref{sec:Calculation}, we provide the details of the calculation. Section \ref{sec:Results} describes the results, and we conclude in Section \ref{sec:Conclusions}.

\section{The perturbative fragmentation function formalism}\label{sec:PFF}
Fragmentation functions were initially introduced to describe the production of hadrons at high $p_T$ \cite{Berman:1971xz}. In analogy to the initial state factorisation theorem, which states that hadron-collision cross sections can be described, up to power corrections, by a convolution of non-perturbative parton distribution functions and perturbative hard scattering matrix elements, the production of hadrons can be factorised into non-perturbative fragmentation functions $D_{i\to h}$, which describe the transition from the parton $i$ to the hadron $h$, and perturbative hard scattering matrix elements:
\begin{equation}
\frac{d\sigma_h}{dE_h}(E_h) =  \sum_i \bigg(D_{i\to h}\otimes \frac{d\sigma_i}{dE_i}\bigg)(E_h) \equiv \sum_i \int_{0}^1\frac{dz}{z} D_{i\to h}(z)\frac{d\sigma_i}{dE_i}\bigg(\frac{E_h}{z}\bigg)\;.
\label{eq:factorized-x-section}
\end{equation}
The presence of collinear singularities implies that fragmentation functions need to be renormalised. This collinear renormalisation is of the form (see also Section \ref{sec:CollRen})
\begin{equation}
D_{i\to h}^B(z) = \sum_{j}\big(Z_{ij}\otimes D_{j\to h}\big)(z)
\end{equation}
and leads to a set of renormalisation group equations for fragmentation functions, called the (time-like) Dokshitzer-Gribov-Lipatov-Altarelli-Parisi (DGLAP) evolution equations \cite{Altarelli:1977zs,Dokshitzer:1977sg,Gribov:1972ri}:
\begin{equation}
\mu_{Fr}^2\frac{dD_{i\to h}}{d\mu_{Fr}^2}(z,\mu_{Fr}) = \sum_j \big(P^{\text{T}}_{ij}\otimes D_{j\to h}\big)(z,\mu_{Fr})\;,
\label{eq:DGLAP}
\end{equation}
where $P^{\text{T}}_{ij}$ are the time-like splitting functions and $\mu_{Fr}$ is the final-state factorisation (or \textit{fragmentation}) scale.

In Ref.\ \cite{Mele:1990cw}, it was pointed out that the same formalism can be used to describe the production of a massive quark starting from a massless calculation. The difference is that the production of a heavy quark can be described perturbatively, and the corresponding fragmentation functions can thus be calculated in perturbation theory. The main benefit of factorising the production of the massive quark into fragmentation functions is that this allows one to use the DGLAP equations to resum mass logarithms: whereas the fully massive calculation contains logarithms of the type $\ln(Q^2/m^2)$, the perturbative fragmentation function (PFF) formalism splits the calculation into fragmentation functions, which contain the logarithm $\ln(\mu_{Fr}^2/m^2)$, and massless cross sections, which contain the logarithm $\ln(Q^2/\mu_{Fr}^2)$. It is thus possible to evaluate the fragmentation functions at a scale $\mu_{Fr}\sim m$ and then use the DGLAP equations to obtain them at a scale $\mu_{Fr}\sim Q$ without introducing any large logarithms. The accuracy of the resummation of mass logarithms is given by the order of the splitting functions used, i.e.\ leading-logarithmic accuracy (LL) for LO splitting functions, NLL for NLO splitting functions, etc.

While the original derivation of the PFFs at NLO in Ref.\ \cite{Mele:1990cw} involved comparing the massive and massless cross sections, the derivation of the NNLO heavy-quark FFs in Refs.\ \cite{Melnikov:2004bm,Mitov:2004du} used a definition equivalent to the field-theoretic definition given in Ref.\ \cite{Collins:1981uw}. This formalism can be used for the production of any heavy particle and was first applied to Higgs-boson production in Ref.\ \cite{Braaten:2015ppa}. There, the LO FF and splitting function for the transition $t\to H$, and of several other transitions, was derived by directly employing the definition of FFs given in Ref.\ \cite{Collins:1981uw}. The fully-differential factorisation formula for this case reads
\begin{equation}
d\sigma_H =  \sum_i D_{i\to H}\otimes d\sigma_i|_{m_H=m_t=0}\;,
\end{equation}
where the sum is over all QCD partons (including the top quark) and the Higgs boson, assuming only QCD corrections are considered.

\section{Details of the calculation}\label{sec:Calculation}
In this section we present our calculation in considerable detail. We set the notation, discuss the renormalisation procedure and, in particular, describe the techniques used to perform the calculation of the Feynman integrals.

\subsection{Notation and definitions}
If $n^\mu$ is an arbitrary four-vector satisfying $n^2=0$ and $n\cdot \overline{n}=1$, where $\overline{n}^\mu = n_\mu$, then any four-vector $p^\mu$ can be decomposed as follows:
\begin{equation}
p^\mu = p^+\overline{n}^\mu+p^-n^\mu+p_T^\mu = (p\cdot n)\overline{n}^\mu+(p\cdot \overline{n})n^\mu+p_T^\mu,\;\;\;\;p_T\cdot n = p_T\cdot\overline{n} = 0,\;\;\;\;p_T^2 < 0\;.
\end{equation}

The bare fragmentation function $D^B_{q\to H}(z)$ for a quark $q$ to fragment to a Higgs boson $H$ with momentum $p_H^\mu$ with $p_H^+ = zp_q^+$ and $p_{H,T}^\mu=0$ is defined as follows \cite{Collins:1981uw}:
\begin{align}
D^B_{q\to H}(z) =\:&\frac{z^{d-3}}{4\pi}\int dx^- e^{-ip_H^+x^-/z}\frac{1}{2N_c}\text{Tr}_\text{color}\text{Tr}_\text{Dirac}\bigg[\slashed{n}\big\langle 0\big\lvert\psi_q(0)\notag\\&\times\overline{\text{P}}\exp\bigg(ig\int_0^{\infty}dy^-\:n\cdot A_a(y^-n)\text{T}^\text{T}_a\bigg)a_H^+(p_H)a_H(p_H)\notag\\&\times\text{P}\exp\bigg(-ig\int_{x^-}^{\infty}dy^-\:n\cdot A_b(y^-n)\text{T}^\text{T}_b\bigg)\overline{\psi}_q(x^-n)\big\rvert 0\big\rangle\bigg]\;,
\label{eq:DefFFq}
\end{align}
where $\text{P}$ ($\overline{\text{P}}$) is the (reverse) path-ordering operator.
Here, 'bare' refers to the need to perform collinear renormalisation, as explained in Section \ref{sec:CollRen}. The bare gluon fragmentation function $D^B_{g\to H}(z)$ is defined as \cite{Collins:1981uw}
\begin{align}
D^B_{g\to H}(z) =\:&\frac{-z^{d-2}}{(N_c^2-1)(d-2)2\pi p_H^+}\int dx^- e^{-ip_H^+x^-/z}\big\langle 0\big\lvert n_{\alpha}F^{\alpha}_{a\mu}(0)\notag\\&\times\bigg[\overline{\text{P}}\exp\bigg(-ig\int_0^{\infty}dy^-\:n\cdot A_d(y^-n)\text{T}_d\bigg)\bigg]_{ab}a_H^+(p_H)a_H(p_H)\notag\\&\times\bigg[\text{P}\exp\bigg(ig\int_{x^-}^{\infty}dy^-\:n\cdot A_e(y^-n)\text{T}_e\bigg)\bigg]_{bc}n_{\beta}F_{c}^{\beta\mu}(x^-n)\big\rvert 0\big\rangle\;.
\label{eq:DefFFg}
\end{align}

Using these definitions, fragmentation functions can be calculated using standard higher-order techniques for the computation of cross sections. We use dimensional regularisation with $d=4-2\epsilon$ dimensions and the $\overline{\text{MS}}$ definition of the renormalisation scale:
\begin{equation}
\overline{\mu}^2=\frac{e^{\gamma_\text{E}}}{4\pi}\mu^2\;.
\end{equation}

At NLO, the top-quark mass (and the Yukawa coupling) as well as the wave-function are renormalised on-shell. We work in the linear gauge with an arbitrary gauge parameter.

The computation of $D^B_{t\to H}$ at $\mathcal{O}(y_t^2)$ using eq.\ \eqref{eq:DefFFq} is simple enough to reproduce here explicitly. There is only a single diagram, shown in fig.\ \ref{fig:LOandGlu}, which yields (cf.\ Ref.\ \cite{Braaten:2015ppa})
\begin{align}
D^B_{t\to H}(z) &= \frac{z^{d-3}}{4\pi}\int \frac{d^dp_t}{(2\pi)^d}(2\pi)\delta^+(p_t^2-m_t^2)(2\pi)\delta(p_H^+/z-(p_t+p_H)^+)\frac{y_t^2\overline{\mu}^{2\epsilon}}{2N_c}\notag\\&\;\;\;\;\times\text{Tr}\bigg[\slashed{n}\frac{\slashed{p}_t+\slashed{p}_H+m_t}{(p_t+p_H)^2-m_t^2}(\slashed{p}_t+m_t)\frac{\slashed{p}_t+\slashed{p}_H+m_t}{(p_t+p_H)^2-m_t^2}\bigg]\notag\\&= \frac{z^{d-5}y_t^2\overline{\mu}^{2\epsilon}}{4\pi}\int \frac{d^{d-2}p_{t,T}}{(2\pi)^{d-2}}\bigg(\frac{(2-z)^2m_t^2}{\big(p_{t,T}^2-(m_t^2+\frac{1-z}{z^2}m_H^2)\big)^2}-\frac{z^2p_{t,T}^2}{\big(p_{t,T}^2-(m_t^2+\frac{1-z}{z^2}m_H^2)\big)^2}\bigg)\notag\\&= \frac{y_t^2}{16\pi^2}z\bigg[\frac{1}{\epsilon}-\text{ln}\bigg(\frac{m_t^2z^2+m_H^2(1-z)}{\mu^2}\bigg)+\bigg(4m_t^2-m_H^2\bigg)\frac{1-z}{m_t^2z^2+m_H^2(1-z)}+\mathcal{O}(\epsilon)\bigg]\;.
\label{eq:LOFF}
\end{align}

Note that the terms $\mathcal{O}(\epsilon)$ of this LO result are required for the different renormalisation contributions to the NLO result.

\subsection{Collinear renormalisation}\label{sec:CollRen}
The parton-to-Higgs FFs are renormalised in full analogy to the more common heavy-quark case, presented through NNLO in e.g.\ Ref.\ \cite{Czakon:2021ohs}. Nonetheless, the collinear renormalisation matrix is reproduced here for the case of Higgs fragmentation. We perform collinear renormalisation in the $\overline{\text{MS}}$ scheme. Setting $\mu_{Fr}=\mu_R$ for convenience, the NLO QCD $\overline{\text{MS}}$ collinear renormalisation of $D_{t\to H}$ is given by
\begin{align}
D_{t\to H}^B(z) &\:= \big(Z_{tH}\otimes D_{H\to H}\big)(z)+\big(Z_{tt}\otimes D_{t\to H}\big)(z)+\sum_{i\neq t,H}\big(Z_{ti}\otimes D_{i\to H}\big)(z)\notag\\&\:=  Z_{tH}(z)+\big(Z_{tt}\otimes D_{t\to H}\big)(z)+\mathcal{O}(y_t^2\alpha_s^2,\:y_t^4)\;,\label{eq:CollRenTop}\\\notag\\
Z_{tH}(z) &\:= \frac{y_t^2}{16\pi^2}\frac{1}{\epsilon}P^{(0)\text{T}}_{tH}(z)\notag\\&\:+\frac{y_t^2}{16\pi^2}\frac{\alpha_s}{2\pi}\bigg(\frac{1}{2\epsilon}P^{(1)\text{T}}_{tH}(z)+\frac{1}{2\epsilon^2}\big(P^{(0)\text{T}}_{qq}\otimes P^{(0)\text{T}}_{tH}\big)(z)-\frac{\beta_t^{(0)}}{4\epsilon^2}P^{(0)\text{T}}_{tH}(z)\bigg)+\mathcal{O}(y_t^2\alpha_s^2,\:y_t^4)\;,\\\notag\\
Z_{tt}(z) &\:= \delta(1-z)+\frac{\alpha_s}{2\pi}\frac{1}{\epsilon}P^{(0)\text{T}}_{qq}(z)+\mathcal{O}(\alpha_s^2,\:y_t^2)\;. 
\end{align}
Here 
\begin{equation}
D_{H\to H}(z) = \delta(1-z)+\mathcal{O}(y_t^2)
\end{equation}
has been used, $P_{qq}^{(0)\textrm{T}}$ is the standard LO time-like quark-to-quark splitting function,
\begin{equation}
P_{qq}^{(0)\textrm{T}}(z) = C_F\left[\frac{1+z^2}{(1-z)_+}+\frac{3}{2}\delta(1-z)\right]\;,
\end{equation}
and $\beta_t^{(0)}$ is defined through
\begin{equation}
\mu_R^2\frac{dy_t^2}{d\mu_R^2} \equiv \beta_t = -\epsilon y_t^2-\frac{y_t^2\alpha_s}{4\pi}\beta_t^{(0)}+\mathcal{O}(y_t^2\alpha_s^2,\:y_t^4)\;,
\end{equation}
which implies
\begin{equation}
\beta_t^{(0)} = 2\gamma_0 = 6C_F\;,
\end{equation}
where $\gamma_0$ is the leading-order quark-mass anomalous dimension.

The NLO QCD $\overline{\text{MS}}$ collinear renormalisation of the gluon-to-Higgs FF is given by
\begin{align}
D_{g\to H}^B(z) &\:= \big(Z_{gg}\otimes D_{g\to H}\big)(z)+\big(Z_{gH}\otimes D_{H\to H}\big)(z)\notag\\&\:+2\big(Z_{gt}\otimes D_{t\to H}\big)(z)+\sum_{i\neq t,H}\big(Z_{ti}\otimes D_{i\to H}\big)(z)\notag\\&\:= D_{g\to H}(z)+Z_{gH}(z)+2\big(Z_{gt}\otimes D_{t\to H}\big)(z)+\mathcal{O}(y_t^2\alpha_s^2,\:y_t^4)\;,\label{eq:gluCollRen}\\\notag\\
Z_{gg}(z) &\:= \delta(1-z)+\mathcal{O}(\alpha_s)\;,\\\notag\\
Z_{gH}(z) &\:= \frac{y_t^2}{16\pi^2}\frac{\alpha_s}{2\pi}\frac{1}{2\epsilon}P^{(1)\text{T}}_{gH}(z)+\mathcal{O}(\alpha_s^2,\:y_t^2)\;,\\\notag\\
Z_{gt}(z) &\:= \frac{\alpha_s}{2\pi}\frac{1}{\epsilon}P^{(0)\text{T}}_{gq}(z)+\mathcal{O}(\alpha_s^2,\:y_t^2)\;. 
\end{align}
The factor 2 in the second line of eq.\ \eqref{eq:gluCollRen} accounts for both top and anti-top quarks, and
\begin{equation}
P^{(0)\text{T}}_{gq}(z) = T_F\big[z^2+(1-z)^2\big]\;
\end{equation}
is the usual LO time-like gluon-to-quark splitting function.

$P^{(0)\text{T}}_{tH}(z)$ was first calculated in Ref.\ \cite{Braaten:2015ppa}. It can also be obtained from the results above: since the poles in $\epsilon$ of the bare FF must be removed by the collinear renormalisation and the only counterterm at LO is proportional to $P^{(0)\text{T}}_{tH}(z)$, a simple comparison of eqs.\ \eqref{eq:LOFF} and \eqref{eq:CollRenTop} yields
\begin{align}
P^{(0)\text{T}}_{tH}(z)=z\;.
\end{align}

This result, together with the well-known QCD splitting functions and the top-Yukawa $\beta$-function, allows us to predict all the poles of the bare Higgs fragmentation functions at NLO, except for the terms proportional to $P^{(1)\text{T}}_{tH}$ and $P^{(1)\text{T}}_{gH}$, since both of these functions were previously unknown. However, similarly to what was done for the LO case above, both of these splitting functions can be readily derived from the corresponding NLO FFs and have thus been obtained as a by-product of the computation presented here.

\subsection{Outline of the calculation and computational techniques}
The general outline of the calculation is identical for both FFs. We start by determining all relevant Feynman diagrams, as shown in Appendix \ref{sec:FeynDia}. All subsequent symbolic manipulations were performed using Mathematica \cite{Mathematica}. The Mathematica package FeynCalc \cite{FeynCalc} was used to perform the contraction of Lorentz indices and Dirac traces. On-shell conditions and the momentum fraction $\delta$-distribution coming from the definition of the fragmentation function were replaced with cut propagators using reverse unitarity \cite{Anastasiou:2002yz,Anastasiou:2002wq,Anastasiou:2002qz,Anastasiou:2003yy,Anastasiou:2003ds}:
\begin{equation}
\delta(x) = -\frac{1}{2\pi i}\frac{1}{(x)_c} \equiv -\frac{1}{2\pi i}\bigg(\frac{1}{x+i\varepsilon}-\frac{1}{x-i\varepsilon}\bigg)\;.
\end{equation}
For each part of the calculation ($D_{g\to H}$, the real corrections to $D_{t\to H}$ and the virtual corrections to $D_{t\to H}$), all required Feynman integrals can be written as the product of 7 propagators raised to integer powers. These integrals are then reduced to a smaller set of master integrals using IBP relations \cite{Tkachov:1981wb,Chetyrkin:1981qh}. This reduction was performed using FIRE6 \cite{Smirnov:2019qkx}.

The master integrals were calculated explicitly in the $m_t\to \infty$ limit (or equivalently $m_H\to 0$) up to the required order in $\epsilon$, making use of HypExp \cite{Huber:2005yg,Huber:2007dx,Maitre:2005uu,Maitre:2007kp} to expand hypergeometric functions in $\epsilon$ where necessary. Subsequently, the method of differential equations \cite{Kotikov:1990kg,Kotikov:1991hm,Bern:1992em,Bern:1993kr,Kotikov:1993zf,Fleischer:1997bw,Gehrmann:1999as} was used to obtain the master integrals for finite values of $m_t$. The set of master integrals was constructed to be in canonical form, significantly simplifying this step \cite{Henn:2013pwa}. The details of the canonical basis are described in Section \ref{sec:CanonicalBasis}. The list of master integrals for the top-to-Higgs FF, as well as their $\epsilon$-expansions through the order required for the present calculation, can be found in Appendix \ref{sec:FeynInts}. The solution of the differential equations was performed using tools provided by PolyLogTools \cite{Duhr:2019tlz}. The results, initially involving multiple polylogarithms (MPLs) of weight three, were simplified using the strategy of symbols \cite{Goncharov.A.B.:2009tja,Goncharov:2010jf,Duhr:2011zq} outlined in detail in Ref.\ \cite{Goncharov:2010jf} and summarised for the present case in Section \ref{sec:symbols}, again making use of tools provided by PolyLogTools. The final results are expressed in terms of a comparatively small number of classical polylogarithms.

The calculation was performed under the assumption that $m_H<\sqrt{2}m_t$. As such, the results presented below are only valid for values of the masses satisfying this condition, which however includes the values relevant for the Standard Model. Of course, the NLO splitting functions are valid for any values of the masses.

\subsection{The canonical basis}\label{sec:CanonicalBasis}
As mentioned in the previous section, the differential equations method was used to derive the top mass dependence of the master integrals. The system of first-order linear differential equations with respect to $m_t^2$ reads 
\begin{equation}
\partial_{m_t^2} \vec{g}(m_t^2, \epsilon) = A(m_t^2,\epsilon)\, \vec{g}(m_t^2, \epsilon)\;,
\label{eq:GenDE}
\end{equation}
where $\vec{g}(m_t^2, \epsilon)$ is the vector of master integrals and $A(m_t^2,\epsilon)$ is a $N \times N$ matrix, with $N$ being the dimension of $\vec{g}(m_t^2, \epsilon)$. Note that for simplicity of notation the dependence on kinematic invariants other than $m_t^2$ is not explicit in eq.\ \eqref{eq:GenDE}. This convention will be used henceforth. 

A modern approach for the solution of  eq.\ \eqref{eq:GenDE} consists in the application of the canonical basis method. For the set of master integrals arising from the reduction of both the real and the virtual contributions to $D_{t\to H}^{(1)}$, a change of basis was found such that
\begin{equation}
\mathrm{d} \vec{f}(m_t^2, \epsilon) = \epsilon\, \mathrm{d} \Tilde{A}(m_t^2)\, \vec{f}(m_t^2, \epsilon)\;,
\label{eq:GenCanDE}
\end{equation}
where $\vec{f}(m_t^2, \epsilon)$ is the vector of canonical master integrals, chosen to be free of $\epsilon$-poles, $\mathrm{d} \Tilde{A}(m_t^2) = \Tilde{A}(m_t^2)\, \mathrm{d} m_t^2$, and the $\epsilon$-dependence is factorised so that $\Tilde{A}(m_t^2)$ depends only on the kinematic invariants. Moreover, we define the canonical basis elements $f_i(m_t^2,\epsilon)$ as the product of the so-called pre-canonical integrals $I_i(m_t^2, \epsilon)$, the algebraic pre-factors $B_i(m_t^2)$, which are functions of the
kinematic invariants, and the dimensional regularisation parameter to the $n_i$-th power:
\begin{equation}
f_i(m_t^2,\epsilon)= \epsilon^{n_i}\, B_i(m_t^2)\, I_i(m_t^2, \epsilon)\;,
\label{eq:GenCanMI}
\end{equation}
with $n_i \in \mathbb{N}$. 

The canonical basis was obtained by the use of the following semi-algorithmic approach:
\begin{itemize}
    \item The pre-canonical integrals $I_i(m_t^2, \epsilon)$ were chosen by modifying the powers of the
    denominators of the master integrals with the aim of maximising their symmetries.
    \item The $n_i$ power of $\epsilon$ was chosen so that the product $\epsilon^{n_i}\, I_i(m_t^2, \epsilon)$ contains no $\epsilon$ poles.
    \item After the steps outlined above, the differential equations are linear in $\epsilon$. At this stage, the Magnus Transformation algorithm (suggested in Ref.\ \cite{DiVita:2014pza}) was applied in order to determine the $B_i(m_t^2)$ pre-factors, leading to a system of differential equations as in eq.\ \eqref{eq:GenCanDE}.
\end{itemize}

The canonical form for the topology of the real corrections to the top-to-Higgs FF (defined in eq.\ \eqref{eq:TopoReal}) was obtained by performing the change of basis:
\begin{equation}
\begin{aligned}
& f_1=\epsilon^2 \ n\cdot p_H \  I_{0,0,0,1,1,2,1}\;,\\
& f_2=\epsilon^2 \ n\cdot p_H \  I_{0,0,0,1,2,1,1}\;,\\
& f_3=\epsilon^2  m_H \sqrt{4 m_t^2-m_H^2} \ n\cdot p_H \  I_{0,0,1,1,1,2,1}\;,\\
& f_4=\epsilon^2 \ m_t^2 \ n\cdot p_H \ I_{0,0,1,2,1,1,1}\;,\\
& f_5=\epsilon^3\ n\cdot p_H \  I_{0,0,1,1,1,1,1}\;,\\
& f_6=\epsilon^2  m_H \sqrt{4 m_t^2-m_H^2} \ n\cdot p_H \ I_{0,1,2,0,1,1,1}\;,\\
& f_7=\epsilon^2 \ \frac{1-z}{z} \  (n\cdot p_H)^2 \  I_{0,1,1,0,1,1,2}\;,\\
& f_8=\epsilon^3 \ n\cdot p_H \  I_{0,1,1,0,1,1,1}\;.
\end{aligned}
\label{eq:CanBasReal}
\end{equation} 
Similarly, the canonical form for the topology of the virtual corrections to the top-to-Higgs FF (defined in eq.\ \eqref{eq:TopoVirt}) was obtained by performing the change of basis:
\begin{equation}
\begin{aligned}
& f_1=\epsilon^2 \ n\cdot p_H \  I_{0,2,1,0,0,1,1}\;,\\
& f_2=\epsilon^2 m_H \sqrt{4 m_t^2-m_H^2} \ n\cdot p_H \  I_{0,2,1,1,0,1,1}\;,\\
& f_3=\epsilon^3 \ (n\cdot p_H)^2 \  I_{1,1,1,1,0,1,1}\;,\\
& f_4=\epsilon^2 \ n\cdot p_H \  I_{0,0,0,1,2,1,1}\;,\\
& f_5=\epsilon^2 \ n\cdot p_H \  I_{0,0,0,2,1,1,1}\;,\\
& f_6=\epsilon^2 \ \frac{1-z}{z}  \ (n\cdot p_H)^2\  I_{0,1,0,1,1,1,2}\;,\\
& f_7=\epsilon^2 m_H \sqrt{4 m_t^2-m_H^2} \ n\cdot p_H \  I_{0,1,0,2,1,1,1}\;,\\
& f_8=\epsilon^3 \ n\cdot p_H \  I_{0,1,0,1,1,1,1}\;.
\end{aligned}
\label{eq:CanBasVirt}
\end{equation}
The Feynman diagram representation of the pre-canonical integrals is given in Appendix \ref{sec:FeynDiaMIs}, while their analytic expressions are provided in Appendix \ref{sec:FeynInts}.

Starting from eq.\ \eqref{eq:GenCanDE}, the change of variable $m_t^2 \to m_H^2(1-\tau^2)/4$ was performed in order to linearise the square-root dependence on $m_t^2$. Hence, the canonical master integrals of both the real and the virtual corrections to the top-to-Higgs FF satisfy the following system of differential equations:
\begin{equation}
    \mathrm{d}\vec{f}(\tau, \epsilon)=\epsilon \, \mathrm{d}\Tilde{A}({\tau}) \, \vec{f}(\tau, \epsilon).
\label{eq:CanDEt}
\end{equation}
The differential equation matrix can be written in $\mathrm{d}\!\log$ form:
\begin{equation}
\mathrm{d}\Tilde{A}({\tau}) = \sum_{i=1}^6 M_i \, \mathrm{d}\!\log\left(\alpha_i\right),
\label{eq:dLogForm}
\end{equation}
where the matrices $M_i$ are $8 \times 8$ matrices with purely rational entries. The arguments $\alpha_i$ and the matrices $M_i$ are given in Appendix \ref{sec:CanMat}. 

The integration of the master integrals was performed order by order in $\epsilon$ in terms of MPLs. The integration constants are $m_H$ and $z$ dependent; therefore the boundary conditions were computed by calculating the master integrals in the $m_H \to 0$ limit. 

\subsection{Simplifying polylogarithmic expressions using symbols}\label{sec:symbols}
As explained in Refs.\  \cite{Goncharov.A.B.:2009tja,Goncharov:2010jf,Duhr:2011zq}, the symbol map is an operation which maps polylogarithmic functions, e.g.\ the classical polylogarithms and MPLs, onto a direct product of weight-1 objects, i.e.\ logarithms. Note that the constants $\pi^n$ and $\zeta(n)$ count as weight-n objects as well. For example, the symbol of a classical polylogarithm is given by\footnote{The symbol $\otimes$ does not represent a convolution in this section, but rather a tensor product.}
\begin{equation}
\text{symbol}(\text{Li}_n(x)) = \text{ln}(1-x)\otimes\underbrace{\text{ln}(x)\otimes...\otimes\text{ln}(x)}_{\text{n-1 times}}\;.\label{eq:symbolPL}
\end{equation}
A few properties of symbols are important for the present discussion. First, the symbol map is linear, i.e.
\begin{equation}
\text{symbol}(a+b) = \text{symbol}(a)+\text{symbol}(b)\;.
\end{equation}
Second, the operator $\otimes$ is distributive, i.e.
\begin{equation}
...\otimes(a+b)\otimes... = (...\otimes a\otimes...)+(...\otimes b\otimes...)\;,
\end{equation}
giving the important relation
\begin{equation}
...\otimes\text{ln}(a\cdot b)\otimes... = (...\otimes\text{ln}(a)\otimes...)+(...\otimes\text{ln}(b)\otimes...)\;.\label{eq:symbolDistr}
\end{equation}
Third, the symbol of a product of two functions is given by the shuffle product of the symbols:
\begin{equation}
\begin{gathered}
\text{symbol}(f) = a_1\otimes...\otimes a_n\;,\;\;\;\;\text{symbol}(g) = a_{n+1}\otimes...\otimes a_{n+m}\;,\\
\text{symbol}(f\cdot g) = \sum_{\sigma\in\Sigma(n,m)} a_{\sigma^{-1}(1)}\otimes...\otimes a_{\sigma^{-1}(n+m)}\;,
\end{gathered}
\end{equation}
where $\Sigma(n,m)$ is the set of all permutations $\sigma$ of $1,...,n+m$ satisfying
\begin{equation}
\sigma(1)<...<\sigma(n)\;,\;\;\;\;\sigma(n+1)<...<\sigma(n+m)\;.
\end{equation}
Another property relevant here is that $\zeta(n)$ and higher powers of $\pi$ are mapped to zero. The final property needed is that if the symbols of two functions are identical, then the functions are identical up to contributions in the kernel of the symbol map, i.e.\ terms containing factors of $\zeta(n)$ or $\pi^n$ for $n>1$.

Using these properties, and with the help of tools implemented in PolyLogTools, one can now simplify the initial form of the master integrals as resulting from the solution of the differential equations via the following steps. First, one takes the symbol of the initial form. This allows one to work with logarithms and their simple relations, i.e.\ eq.\ \eqref{eq:symbolDistr}, reducing the number of distinct logarithms to a smaller set with simpler arguments. One then searches for a linear combination of classical polylogarithms with simple arguments with the same symbol. This linear combination must be identical to the initial form of the master up to terms in the kernel of the symbol map, of which there are only a few at low weights and whose numerical coefficients are rational numbers. These terms can thus be obtained by fitting their coefficients over the rational numbers. Finally, a high-precision numerical comparison of the initial form and the simplified one over the entire relevant region of the parameter space\footnote{Here, the parameters are $z$, $m_H$ and $m_t$.} is used to confirm the fitted coefficients. Note that it is well known that through weight three, all MPLs can be written as linear combinations of classical polylogarithms, while this is not the case for higher weights. Since the present problem does not contain higher-weight MPLs, classical polylogarithms will always be sufficient.

Finding a function with the same symbol is the most laborious part of the procedure above. Fortunately, as explained in Ref.\ \cite{Goncharov:2010jf}, different types of contributions satisfy different symmetry relations with respect to the various entries of the symbol tensor, allowing one to isolate a specific type of contribution, find its simplified form and subtract it from the full master, after which one can move on to the next type of contribution. For example, at weight two, there are four types of contributions:
\begin{equation}
\text{Li}_2(a)\;,\;\;\;\;\text{ln}(a)\text{ln}(b)\;,\;\;\;\;\pi\:\text{ln}(a)\;,\;\;\;\;\pi^2\;.
\end{equation}
Initially ignoring contributions containing factors of $\pi$ leaves $\text{Li}_2(a)$ and $\text{ln}(a)\text{ln}(b)$. The symbol of the latter is symmetric with respect to the first and the second entry of the symbol:
\begin{equation}
\text{symbol}(\text{ln}(a)\text{ln}(b)) = \text{ln}(a)\otimes\text{ln}(b)+\text{ln}(b)\otimes\text{ln}(a)\;,
\end{equation}
whereas the symbol of $\text{Li}_2(a)$ is not (cf.\ eq.\ \eqref{eq:symbolPL}). Anti-symmetrising with respect to these entries thus only leaves a part given by dilogarithmic contributions, which can now be determined more easily. After subtracting the dilogarithmic part, all contributions to the symbol are of the form $\text{ln}(a)\text{ln}(b)$ or $\pi\:\text{ln}(a)$ and can essentially be directly read off from the symbol. The reason why the contributions involving $\pi$ are initially ignored is that $\pi$ behaves in an exceptional way within the symbol, making (anti-)symmetrisation impossible. The only contributions left must be proportional to $\pi^2$ and can thus be trivially determined via a numerical comparison.

At weight three, there are 8 contributions:
\begin{equation}
\begin{gathered}
\text{Li}_3(a)\;,\;\;\;\;\text{ln}(a)\text{Li}_2(b)\;,\;\;\;\;\pi\:\text{Li}_2(a)\;,\;\;\;\;\text{ln}(a)\text{ln}(b)\text{ln}(c)\;,\\\pi\:\text{ln}(a)\text{ln}(b)\;,\;\;\;\;\pi^2\:\text{ln}(a)\;,\;\;\;\;\pi^3\;,\;\;\;\;\zeta(3)\;.
\end{gathered}
\end{equation}
Anti-symmetrising with respect to the last two entries of the symbol leaves only $\text{ln}(a)\text{Li}_2(b)$ and $\pi\:\text{Li}_2(a)$, with the first entry determined by the logarithms and the last two entries by the dilogarithms. After subtracting these contributions from the full result, anti-symmetrisation with respect to the first two entries of the symbol (ignoring again terms containing $\pi$) leaves only the trilogarithms. After those have been determined and subtracted, only the purely logarithmic terms remain and can be read off directly. The terms of the type $\pi^2\:\text{ln}(a)$, $\pi^3$ and $\zeta(3)$ must again be fitted. Here, the knowledge of which logarithms showed up in the simplification so far is used as a guide for the Ansatz: indeed, no logarithms were needed for these fits that were not already needed for the contributions that can be determined using the symbol. In theory, terms proportional to $\pi^3$ or $\zeta(3)$ can only be separated by requiring their rational coefficients to be reasonable, after which the high-precision comparison of the initial and final forms is used to confirm their correctness. In practice, at least for the present computation, the $\pi^3$-terms are always accompanied by the imaginary unit $i$, while the $\zeta(3)$-terms are not, allowing for a much simpler separation.

\section{Results}\label{sec:Results}

The non-zero NLO parton-to-Higgs splitting functions are
\begin{align}
P^{(1)\text{T}}_{tH}(z) &\:= C_F\bigg[-8z\:\text{Li}_2(z)+z\ln^2(1-z)-\frac{1}{2}z\ln^2(z)-4z\ln(z)\ln(1-z)\notag\\&\:\phantom{{}=C_F\bigg[}+3z\ln(1-z)-\bigg(1-\frac{1}{2}z\bigg)\ln(z)-\bigg(\frac{13}{2}-15z\bigg)\bigg]\;,\notag\\
P^{(1)\text{T}}_{gH}(z) &\:= 2T_F\big[z\:\text{ln}^2z-(1+5z)\:\text{ln}\:z-(6-4z-2z^2)\big]\;.
\end{align}
As required, they are independent of $m_t$ and $m_H$, which is an important test of their validity. They have been extracted from the full calculation, as well as from simplified calculations where either $m_t$ or $m_H$ is set to zero. The results agree in all three cases, providing a further cross-check of the results. 

\medskip

Defining the shorthand notation
\begin{align}
L &\:= \ln\bigg(\frac{m_t^2}{\mu^2}\bigg)\;,\;\;\;\;\xi = \frac{m_H}{2m_t}\;,\;\;\;\;\beta = \sqrt{1-\xi^2}\;,\;\;\;\;r = \frac{z^2}{z^2+4(1-z)\xi^2}\;,\;\;\;\;0<r<1\;,\notag\\t &\:= \frac{4 \xi ^2 (1-z)}{z+4(1-z)\xi ^2}\;,\;\;\;\;0<t<1\;,\;\;\;\;x^\pm = 2\xi(\xi\pm i\beta)\;,\;\;\;\;w^\pm = \frac{2 \xi  (1-z) (\xi  (2-z)\mp i \beta  z)}{z^2+4(1-z)\xi^2}\;,\notag\\y &\:= \frac{2 \xi}{z}(\sqrt{(1-z)(z+(1-z)\xi^2)}-(1-z)\xi)\;,\;\;\;\;0<y<1\;,\;\;\;\;\;v^\pm = -\frac{1-z}{z}x^\pm\;,\;\;\;\;
\end{align}
the NLO corrections to the fragmentation functions read
\begin{align}
D_{g\to H}^{(1)}(z)&\:=\frac{\alpha_sT_F}{2\pi}\frac{y_t^2}{16\pi^2}\bigg\{ \bigg[2 z \ln (z)-2 z^2+z+1\bigg]L^2+\bigg[-4 z \text{Li}_2(1-r)-8 z \Re\big[\text{Li}_2(v^+)\big]\notag\\&\: -2 z \ln ^2(r)+2 z \ln ^2(z)+\bigg(-\frac{2 z}{\xi ^2}-8 \xi ^2+2 z^2+4\bigg)\ln (r) +\frac{\beta}{\xi ^3}\bigg(8\xi ^4-2\xi ^2 (1+z-z^2)\notag\\&\:+ z^2\bigg) \arg \bigg(\frac{1-v^+}{1-v^-}\bigg)+2 \bigg(8 \xi ^2+7 z-3\bigg) \ln (z)+4 (z-1) (\frac{z}{\xi ^2}-2z-2)\bigg]L\notag\\&\:+32 z \Re\bigg[\text{Li}_3\bigg(-\frac{z v^+}{1-z}\bigg)\bigg]+16 z \Re\bigg[\text{Li}_3\bigg(\frac{z v^+}{1-z}+1\bigg)\bigg]+16 z \Re\bigg[\text{Li}_3\bigg(\frac{v^+ (y-1)}{y}\bigg)\bigg]\notag\\&\:+8 z \text{Li}_3(1-y)+16 z \Re\bigg[\text{Li}_3\bigg(\frac{y}{v^+}\bigg)\bigg]-4 z \text{Li}_3\bigg(1-\frac{1}{z}\bigg)-\frac{4}{3 \xi ^3 \beta} \bigg(6 \xi ^4 z^2+\xi ^2 (5-3 z^2)\notag\\&\:-3 z^2\bigg) \Im\big[\text{Li}_2(2-v^+)\big]-4 \bigg(4 \xi ^2+z^2+3 z-2\bigg)\text{Li}_2(1-r)+\frac{8 z}{3} \bigg(-\frac{z^2}{\xi ^4}+\frac{z^2+3}{\xi ^2}\notag\\&\:+3\bigg) \Re\big[\text{Li}_2(2-v^+)]+\frac{4}{3 \xi ^3 \beta}\bigg(6 \xi ^4 z^2+\xi ^2 (5-3 z^2)-3 z^2\bigg) \Im\bigg[\text{Li}_2\bigg(-\frac{z v^+}{1-z}\bigg)\bigg]\notag\\&\:-\frac{4}{3 \xi ^3 \beta}\bigg(6 \xi ^4 z^2+\xi ^2 (5-3 z^2)-3 z^2\bigg) \Im\bigg[\text{Li}_2\bigg(\frac{v^+}{(1-z) (v^+-1)}\bigg)\bigg]+\frac{4}{3 \xi ^3 \beta} \bigg(6 \xi ^4 (2 z^2\notag\\&\:-z-5)+\xi ^2 (-6 z^2+6 z+11)+24 \xi ^6-6 z^2\bigg) \Im\bigg[\text{Li}_2\bigg(\frac{v^+ (1-y)}{y}+1\bigg)\bigg]\notag\\&\:-\frac{4}{3 \xi ^3 \beta} \bigg(6 \xi ^4 (2 z^2-z-5)+\xi ^2 (-6 z^2+6 z+11)+24 \xi ^6-6 z^2\bigg) \Im\bigg[\text{Li}_2\bigg(1-\frac{y}{v^+}\bigg)\bigg]\notag\\&\:+ \bigg(-16 z \ln (1-y)-8 (4 \xi ^2+z^2+3 z-2)\bigg)\Re\big[\text{Li}_2(v^+)\big]+4 z  \ln (1-y)\text{Li}_2(z)\notag\\&\:-16 z  \ln (1-y)\Re\big[\text{Li}_2((1-v^+) (1-y))\big]+16 z  \ln (1-y)\Re\bigg[\text{Li}_2\bigg(-\frac{(1-v^+) y}{v^+}\bigg)\bigg]\notag\\&\:-8 z \ln (1-y) \text{Li}_2(y)-5 z \ln ^3(1-y)-\frac{1}{3} z \ln ^3(1-z)+3 z \ln (1-y) \ln ^2(1-z)\notag\\&\:+5 z \ln (1-z) \arg ^2\bigg(\frac{v^+}{v^-}\bigg)-5 z \ln (1-z) \ln ^2(1-y)-4 z \ln (1-r) \arg ^2\bigg(\frac{v^+}{v^-}\bigg)\notag\\&\:-9 z \ln (z) \arg ^2\bigg(\frac{v^+}{v^-}\bigg)+4 z \ln (r) \arg ^2\bigg(\frac{v^+}{v^-}\bigg)+z \ln (z) \ln ^2(1-y)+6 z\ln (r) \ln ^2(1-y)\notag\\&\:-6 z \ln (1-r) \ln ^2(1-y)+z \ln (1-y) \ln ^2(z)-4 z \ln (1-y) \ln ^2(r)+z \ln (z) \ln ^2(1-z)\notag\\&\:-3 z \ln (1-y) \arg ^2\bigg(\frac{v^+}{v^-}\bigg)-z \ln (1-z) \ln ^2(z)-2 z \ln (1-y) \ln (1-z) \ln (z)\notag\\&\:+4 z \ln (r) \ln (1-y) \ln (1-z)+4 z \ln (1-r) \ln (r) \ln (1-y)\notag\\&\:+4 z \arg \bigg(\frac{1-v^+}{1-v^-}\bigg) \ln (1-y) \arg \bigg(\frac{v^+}{v^-}\bigg)-2 \bigg(4 \xi ^2+z^2+3 z-2\bigg)\ln ^2(r) +\bigg(-\frac{2 z^3}{3 \xi ^4}\notag\\&\:+\frac{2 (z^2+6) z}{3 \xi ^2}+8 \xi ^2+z-5\bigg)\ln ^2(1-y)+\bigg(8 \xi ^2+2 z^2+\frac{17 z}{2}-\frac{7}{2}\bigg) \ln ^2(z)\notag\\&\:+\frac{1}{3 \xi ^3 \beta}\bigg(6 \xi ^4 z^2+\xi ^2 (5-3 z^2)-3z^2\bigg)\ln (r) \arg \bigg(\frac{1-v^+}{1-v^-}\bigg)-\bigg(\frac{1}{3 \xi ^3 \beta}\big(6 \xi ^4 (2 z^2-z-5)\notag\\&\:+\xi ^2 (-6 z^2+6 z+11)+24 \xi ^6-6 z^2\big)+8 \pi  z\bigg) \ln (1-z)\arg \bigg(\frac{v^+}{v^-}\bigg)-\bigg(\frac{1}{\xi ^3 \beta}\big(-2 \xi ^4 (2 z^2\notag\\&\:+z+5)+\xi ^2 (2 z^2+2 z-3)+8 \xi ^6+2 z^2\big)-16\pi  z\bigg) \ln (z)\arg \bigg(\frac{v^+}{v^-}\bigg) -\bigg(\frac{2}{3 \xi ^3 \beta}\big(6 \xi ^4 z^2\notag\\&\:+\xi ^2 (5-3 z^2)-3 z^2\big)+8\pi  z\bigg)\ln (r) \arg \bigg(\frac{v^+}{v^-}\bigg)+\bigg(\frac{1}{3 \xi ^3 \beta}\big(6 \xi ^4 (2 z^2-z-5)+\xi ^2 (-6 z^2\notag\\&\:+6 z+11)+24 \xi ^6-6 z^2\big)+ 8\pi  z\bigg)\ln (1-y)\arg \bigg(\frac{v^+}{v^-}\bigg)-8 \pi  z \ln (1-y) \arg \bigg(\frac{1-v^+}{1-v^-}\bigg)\notag\\&\:+8\pi  z \ln (1-r)  \arg \bigg(\frac{v^+}{v^-}\bigg)-4 \pi ^2 z \ln (1-r)+\frac{\beta}{3 \xi ^3}\bigg(8 \xi ^4 (z^3-3 z^2+6 z+3)-4 \xi ^2 z (2 z^2\notag\\&\:-12 z+9)+7 z^2\bigg)\arg \bigg(\frac{1-v^+}{1-v^-}\bigg)+\bigg(\frac{2 \pi}{3 \xi ^3 \beta}\big(6 \xi ^4 z^2+\xi ^2 (5-3 z^2)-3 z^2\big)+\frac{8}{3} \xi ^2 (z^3\notag\\&\:-3 z^2+6 z-11)+4 \big(-z^3+5 z^2+(\pi ^2-6) z+3\big)+\frac{1}{3\xi ^2} z (3 z^2-6 z-10)\notag\\&\:+\frac{z^3}{3\xi^4}\bigg)\ln (r)  +\frac{8\pi  z }{3 \xi ^4}\bigg(\xi ^2 (z^2+3)+3 \xi ^4-z^2\bigg)\Im\big[\ln (2-v^+)\big]+\frac{4 \pi  }{3 \xi ^3 \beta}\bigg(6 \xi ^4 z^2+\xi ^2 (5\notag\\&\:-3 z^2)-3 z^2\bigg)\Re\big[\ln (2-v^+)\big]+\bigg(\frac{2 \pi}{3 \xi ^3 \beta}  \big(6 \xi ^4 (2 z^2-z-5)+\xi ^2 (-6 z^2+6 z+11)\notag\\&\:+24 \xi ^6-6 z^2\big)+\frac{8 \pi ^2 z}{3}\bigg) \ln (1-z)+\bigg(-\frac{4 \pi}{3 \xi ^3 \beta}  \big(6 \xi ^4 z^2+\xi ^2 (5-3 z^2)-3 z^2\big)-\frac{2}{3} (-56 \xi ^2\notag\\&\:+18 z^2+10 \pi ^2 z+15)\bigg) \ln (z)-\frac{2}{3}\bigg(8 \xi ^2 (z^3-3 z^2+6 z-4)+(-4 (y+2) z^3+8 (y\notag\\&\:+5) z^2-8 (2 y+7) z+5 \pi ^2 z+24)+4 \frac{1}{\xi ^2} (z-1) z (y (z-3)+2)-\frac{4 y z^2}{\xi^4}\bigg)\ln (1-y)\notag\\&\:-\frac{2}{3} \bigg(\big(-35 z^2+z (36 \zeta (3)+17+3 \pi ^2)+18\big)+\frac{1}{\xi ^2} z \big(\pi ^2 (z^2+3)-4 z+4\big)-\frac{\pi ^2 z^3}{\xi^4}\bigg)\bigg\}\;,
\end{align}
\begin{align}
D_{t\to H}^{(1)}(z)&\:=\frac{\alpha_sC_F}{2\pi}\frac{y_t^2}{16\pi^2}\bigg\{ \bigg[\frac{1}{4} (2-5 z)+z \ln (1-z)-\frac{z}{2} \ln (z)\bigg]L^2+ \bigg[6 z \Re\bigg[\text{Li}_2\bigg(\frac{z}{x^+}\bigg)\bigg]\notag\\&\:+2 z \text{Li}_2(r)+4 z \text{Li}_2(z)+4 z \arg \bigg(\frac{x^+}{x^-}\bigg) \Im\bigg[\ln \bigg(1-\frac{z-x^+}{z-x^-}\bigg)\bigg]-\frac{5 z}{4} \arg^2 \bigg(\frac{x^+}{x^-}\bigg)\notag\\&\:-2 z\Im^2\bigg[\ln \bigg(1-\frac{z-x^+}{z-x^-}\bigg)\bigg]+\frac{3 z}{4} \ln ^2(1-r)+\frac{z}{4} \ln ^2(r)-\frac{3 z}{2} \ln (1-r) \ln (1-z)\notag\\&\:-\frac{z}{2} \ln (r) \ln (1-z)+\frac{z}{2} \ln (1-r) \ln (r)-\frac{z}{4} \ln ^2(1-z)-\frac{z}{2} \ln ^2(z)+4 z \ln (1-z) \ln (z)\notag\\&\:-\frac{r}{z} \bigg(4 \xi ^2 (1-z)+8-8 z+3 z^2\bigg) \ln (1-z)- \frac{r}{\xi  z^2}\bigg(8 \pi  \xi ^3 (1-z) z+2 \pi  \xi  z^3\notag\\&\:+\beta  \big(32 \xi ^4 (1-z)-4 \xi ^2 (2-3 z-z^2)+(2-z) z^2\big)\bigg)\bigg(\Im\bigg[\ln \bigg(1-\frac{z-x^+}{z-x^-}\bigg)\bigg]\notag\\&\:-\arg \bigg(\frac{x^+}{x^-}\bigg)\bigg)- \frac{r}{2 \xi ^2 z^2} \bigg(32 \xi ^6 (1-z)-8 \xi ^4 (2-3 z)-4 \xi ^2 z+z^3\bigg)\ln (r)+\bigg(8 \xi ^2-3\notag\\&\:+\frac{z}{2}\bigg) \ln (z)+ \frac{r}{6 \xi  z^2}\bigg(4 \xi ^3 (1-z) \big(27-(45+4 \pi ^2) z\big)-\xi  z \big(36-63 z+(54+4 \pi ^2) z^2\big)\notag\\&\:-3 \beta  \pi  \big(32 \xi ^4 (1-z)-4 \xi ^2 (2-3 z-z^2)+(2-z) z^2\big)\bigg)\bigg]L\notag\\&\:-\frac{2 \xi ^2 z-(2+z)}{\beta  \xi } \bigg(2\Im[\text{Li}_3(x^+)]-\Im\bigg[\text{Li}_3\bigg(\frac{z}{x^+}\bigg)\bigg]-2\Im\bigg[\text{Li}_3\bigg(\frac{z-1}{z-x^+}\bigg)\bigg]\notag\\&\:-\Im\bigg[\text{Li}_3\bigg(\frac{z}{z-x^+}\bigg)\bigg]-\Im\bigg[\text{Li}_3\bigg(-\frac{(1-z) x^+}{z-x^+}\bigg)\bigg]-\Im\bigg[\text{Li}_3\bigg(1-\frac{1-x^+}{1-t}\bigg)\bigg]\notag\\&\:-\Im\bigg[\text{Li}_3\bigg(1-\frac{x^+}{t}\bigg)\bigg]-\Im\bigg[\text{Li}_3\bigg(1-\frac{z}{1-w^+}\bigg)\bigg]+\frac{1}{2}\Im\bigg[\text{Li}_3\bigg(\frac{r+z (w^+-1)}{z w^+}\bigg)\bigg]\notag\\&\:+\frac{1}{2} \Im\bigg[\text{Li}_3\bigg(\frac{z w^+}{r+z (w^+-1)}\bigg)\bigg]-\frac{1}{2} \Im\bigg[\text{Li}_3\bigg(\frac{r w^+}{r+z(w^+-1)}\bigg)\bigg]\notag\\&\:-\frac{1}{2} \Im\bigg[\text{Li}_3\bigg(\frac{r+z(w^+-1)}{r w^+}\bigg)\bigg]\bigg)-9 z\text{Li}_3(1-z)-2 z \text{Li}_3\bigg(1-\frac{z}{r}\bigg)-4 z \text{Li}_3(1-r)\notag\\&\:-\frac{3 z}{2} \text{Li}_3(r)+2 z \text{Li}_3\bigg(\frac{(z-1) r}{z}\bigg)-2 z \text{Li}_3\bigg(1-\frac{r}{z}\bigg)-10 z \Re\bigg[\text{Li}_3\bigg(\frac{z-1}{z-x^+}\bigg)\bigg]\notag\\&\:-16 z \Re\bigg[\text{Li}_3\bigg(\frac{z}{z-x^+}\bigg)\bigg]-8 z \Re\bigg[\text{Li}_3\bigg(\frac{z (1-x^+)}{z-x^+}\bigg)\bigg]-20 z \Re\bigg[\text{Li}_3\bigg(\frac{z}{x^+}\bigg)\bigg]\notag\\&\:-2 z \Re\bigg[\text{Li}_3\bigg(-\frac{(1-z) x^+}{z-x^+}\bigg)\bigg]+10 z \Re[\text{Li}_3(x^+)]+10 z \Re\bigg[\text{Li}_3\bigg(1-\frac{x^+}{x^-}\bigg)\bigg]\notag\\&\:+\frac{5 z}{2} \Re\bigg[\text{Li}_3\bigg(\frac{z-x^+}{z-x^-}\bigg)\bigg]-2 z \Re\bigg[\text{Li}_3\bigg(1-\frac{1-x^+}{1-t}\bigg)\bigg]+2 z \Re\bigg[\text{Li}_3\bigg(1-\frac{x^+}{t}\bigg)\bigg]\notag\\&\:+\frac{5 z}{2} \Re\bigg[\text{Li}_3\bigg(\frac{(z-x^+) (1-x^-)}{(1-x^+) (z-x^-)}\bigg)\bigg]+8 z \Re[\text{Li}_3(1-w^+)]+16 z \Re[\text{Li}_3(w^+)]\notag\\&\:-5 z \Re\bigg[\text{Li}_3\bigg(\frac{w^+}{w^-}\bigg)\bigg]+ \bigg(4 z \Im\bigg[\ln \bigg(1-\frac{z-x^+}{z-x^-}\bigg)\bigg]-\frac{2 \xi ^2 z-(2+z)}{\beta  \xi } \big(2\ln (r)\notag\\&\:+2\ln (1-z)-3\ln (z)\big)-2\pi  z\bigg)\Im\bigg[\text{Li}_2\bigg(\frac{z-1}{z-x^+}\bigg)\bigg]+\frac{2 \xi ^2 z-(2+z)}{\beta  \xi }\bigg(\ln (1-z)\notag\\&\:-\ln (z)+\ln (r)\bigg)\Im\bigg[\text{Li}_2\bigg(\frac{t}{x^+}\bigg)\bigg]+\frac{2 \xi ^2 z-(2+z)}{\beta  \xi }\bigg(\ln (z)-\ln (r)\bigg)\Im\bigg[\text{Li}_2\bigg(\frac{1-\frac{r}{z}}{w^+}\bigg)\bigg]\notag\\&\: + \bigg( 8 z\arg \bigg(\frac{x^+}{x^-}\bigg)-4 z \Im\bigg[\ln \bigg(1-\frac{z-x^+}{z-x^-}\bigg)\bigg]-\frac{2 \xi ^2 z-(2+z)}{\beta  \xi } \big(\ln (1-z)\notag\\&\:-\ln (z)+\ln (r)\big)-\frac{2 r}{\xi  z^2}\bigg(3 \pi  \xi  z \big(4 \xi ^2 (1-z)+z^2\big)-\beta  \big(32 \xi ^4 (1-z)-4 \xi ^2 (2-3 z-z^2)\notag\\&\:+(2-z) z^2\big)\bigg) \bigg)\Im\bigg[\text{Li}_2\bigg(\frac{z}{x^+}\bigg)\bigg]+\bigg(-\frac{2 \xi ^2 z-(2+z)}{\beta  \xi }\big(\ln (1-z)-2\ln (z)+\ln (r)\big)\notag\\&\:+4 z \Im\bigg[\ln \bigg(1-\frac{z-x^+}{z-x^-}\bigg)\bigg]-\frac{2 r}{\xi  z^2}\bigg(\pi  \xi  z \bigg(4 \xi ^2 (1-z)+z^2\bigg)-\beta  \big(32 \xi ^4 (1-z)\notag\\&\:-4 \xi ^2 (2-3 z-z^2)+(2-z) z^2\big)\bigg)\bigg)\Im[\text{Li}_2(x^+)] -\bigg(z \ln (1-z)+\frac{r}{\xi ^2 z^2}\big(32 \xi ^6 (1-z)\notag\\&\:-4 \xi ^4 (8-5 z-5 z^2)+\xi ^2 (16+12 z-20 z^2-3 z^3)-(4-3 z) z^2\big) \bigg) \text{Li}_2(z)\notag\\&\:+\bigg(4 z \ln (z)-2 z \ln (r)+\frac{r}{\xi ^2 z^2}\big(32 \xi ^6 (1-z)-4 \xi ^4 (4-5 z-z^2)+\xi ^2 (8-3 z^3)\notag\\&\:-(4-3 z) z^2\big)\bigg) \text{Li}_2(r)- \frac{r}{\xi ^2 z^2}\bigg(64 \xi ^6 (1-z)-4 \xi ^4 (12-13 z-3 z^2)-\xi ^2 z (8+8 z-z^2)\notag\\&\:+(2-z) z^2\bigg) \text{Li}_2\bigg(\frac{(z-1) r}{z}\bigg)+\bigg(-z \ln (z)+z \ln (r)+\frac{r}{2 \xi ^2 z^2}\big(64 \xi ^6 (1-z)\notag\\&\:-4 \xi ^4 (2-z-5 z^2)+\xi ^2 z (8-2 z-5 z^2)-4 (1-z) z^2\big)\bigg) \text{Li}_2\bigg(1-\frac{r}{z}\bigg)+\bigg(2 z\ln (1-z)\notag\\&\:-2 z \ln (z)+2 z \ln (r)\bigg) \Re\bigg[\text{Li}_2\bigg(\frac{t}{x^+}\bigg)\bigg]+\bigg(2 z \ln (1-z)+14 z \ln (z)-8 z \ln (r)\notag\\&\:- \frac{2 r}{\xi ^2 z^2}\big(16 \xi ^4 (1-z)-\xi ^2 (16+4 z-6 z^3)+(8-5 z)z^2\big)\bigg) \Re\bigg[\text{Li}_2\bigg(\frac{z}{x^+}\bigg)\bigg]\notag\\&\:- \frac{1}{8 \beta \xi } \bigg(6 \xi ^2 z-(6+3 z)\bigg)\arg^3 \bigg(\frac{x^+}{x^-}\bigg)+\frac{3 z}{8} \ln ^3(1-r)+\frac{13 z}{24} \ln ^3(r)-\frac{z}{24} \ln ^3(1-z)\notag\\&\:+\frac{z}{4} \ln ^3(z)+\frac{5(2 \xi ^2 z-(2+z))}{6 \beta  \xi } \Im^3\bigg[\ln \bigg(1-\frac{z-x^+}{z-x^-}\bigg)\bigg]-\frac{25 z}{8} \arg^2 \bigg(\frac{x^+}{x^-}\bigg) \ln (1-z)\notag\\&\:-\frac{13 z}{4} \arg^2 \bigg(\frac{x^+}{x^-}\bigg) \ln (z)+\frac{5 z}{8} \arg^2 \bigg(\frac{x^+}{x^-}\bigg) \ln (1-r)+\frac{3 z}{2} \arg^2 \bigg(\frac{x^+}{x^-}\bigg) \ln (r)\notag\\&\:-\frac{z}{8} \ln ^2(1-r) \ln (1-z)+\frac{3 z}{4} \ln ^2(1-r) \ln (z)-\frac{3 z}{2} \ln ^2(r) \ln (z)-z \ln ^2(r) \ln (1-z)\notag\\&\:-\frac{3 z}{2} \ln (1-r) \ln ^2(r)+z \ln ^2(r) \ln (t)+z \ln ^2(t) \ln (1-z)-z \ln ^2(t) \ln (z)\notag\\&\:+z \ln (r) \ln ^2(t)-2 z \ln (1-r) \ln ^2(z)+z \ln (r) \ln ^2(z)-z \ln (1-z) \ln ^2(z)\notag\\&\:+2 z \ln (t) \ln ^2(z)-\frac{5 z}{4} \ln ^2(1-r) \ln (r)-\frac{5 z}{4} \ln (r) \ln ^2(1-z)+\frac{5 z}{8} \ln (1-r) \ln ^2(1-z)\notag\\&\:-\frac{z}{4} \ln ^2(1-z) \ln (z)-\frac{5 z}{2} \Re\bigg[\ln \bigg(1-\frac{z-x^+}{z-x^-}\bigg)\bigg] \arg^2 \bigg(\frac{x^+}{x^-}\bigg)-\bigg(\frac{7 z}{2} \ln (1-z)\notag\\&\:+z \ln (z) +\frac{z}{2} \ln (1-r) -\frac{7 z}{2} \ln (r)\bigg) \Im^2\bigg[\ln \bigg(1-\frac{z-x^+}{z-x^-}\bigg)\bigg]\notag\\&\:+ \frac{2 \xi ^2 z-(2+z)}{8 \beta  \xi }\bigg(\bigg(\ln ^2(1-r)-8\Im^2\bigg[\ln \bigg(1-\frac{z-x^+}{z-x^-}\bigg)\bigg]+9\ln ^2(r)+7\ln ^2(z)\notag\\&\:-\ln ^2(1-z)\bigg)\arg \bigg(\frac{x^+}{x^-}\bigg)+\bigg(5\arg^2 \bigg(\frac{x^+}{x^-}\bigg)+3\ln ^2(1-z)+6\ln ^2(z)-\ln ^2(1-r)\notag\\&\:-4\ln ^2(r)\bigg)\Im\bigg[\ln \bigg(1-\frac{z-x^+}{z-x^-}\bigg)\bigg]\bigg)+ \bigg(8 z \ln (z)+ \frac{9 z}{2} \ln (1-z)+ \frac{5 z}{2} \ln (1-r)\notag\\&\:-6 z \ln (r)\bigg)\arg \bigg(\frac{x^+}{x^-}\bigg)\Im\bigg[\ln \bigg(1-\frac{z-x^+}{z-x^-}\bigg)\bigg]-\frac{2 \xi ^2 z-(2+z)}{8 \beta  \xi }\bigg(\bigg(8\ln (t)\ln (z)\notag\\&\:+7\ln (1-z) \ln (z)-11\ln (1-r)\ln (z)-8\ln (r)\ln (1-z)+16\ln (r)\ln (z)\notag\\&\:+6\ln (1-r)\ln (1-z)+10\ln (1-r) \ln (r)-8\ln (t)\ln (1-z)-8\ln (r) \ln (t)\notag\\&\:-2\ln (z) \ln \bigg(\bigg(1-\frac{r}{z}\bigg)^2\bigg)+2\ln (r)\ln \bigg(\bigg(1-\frac{r}{z}\bigg)^2\bigg)\bigg)\arg \bigg(\frac{x^+}{x^-}\bigg)+\bigg(8\ln (r) \ln (t)\notag\\&\:+10\ln (z) \ln (1-z)-6\ln (1-r) \ln (1-z)-4\ln (r) \ln (1-z)-12\ln (1-r) \ln (r)\notag\\&\:-8\ln (z) \ln (t)+14\ln (z) \ln (1-r)+8\ln (t) \ln (1-z)+4\ln (r) \ln \bigg(\bigg(1-\frac{r}{z}\bigg)^2\bigg)\notag\\&\:-4\ln (z) \ln \bigg(\bigg(1-\frac{r}{z}\bigg)^2\bigg)\bigg)\Im\bigg[\ln \bigg(1-\frac{z-x^+}{z-x^-}\bigg)\bigg]\bigg)-\frac{3 z}{2} \ln (1-r) \ln (1-z) \ln (z)\notag\\&\:+\frac{5 z}{2} \ln (r) \ln (1-z) \ln (z)+\frac{7 z}{2} \ln (1-r) \ln (r) \ln (z)-2 z \ln (t) \ln (1-z) \ln (z)\notag\\&\:+2 z \ln (1-r) \ln (t) \ln (z)-2 z \ln (1-r) \ln (r) \ln (t)+3 z \ln (1-r) \ln (r) \ln (1-z)\notag\\&\:-2 z \ln (1-r) \ln (t) \ln (1-z)+z \ln (r) \ln (t) \ln (1-z)-3 z \ln (r) \ln (t) \ln (z)\notag\\&\:+ \frac{r}{16 \beta ^2 \xi ^2 z^2}\bigg(512 \xi ^8 (1-z)-\xi ^6 (832-960 z)-24 \beta  \pi  \xi ^5 (-1+z) z-3 \beta  \pi  \xi  z^2 (2+z)\notag\\&\:-4 z^2 (-8+7 z)+6 \beta  \pi  \xi ^3 (-4+2 z+2 z^2+z^3)-8 \xi ^4 (-56+70 z-4 z^2+3 z^3)\notag\\&\:+4 \xi ^2 (-32+28 z-16 z^2+13 z^3)\bigg) \arg^2 \bigg(\frac{x^+}{x^-}\bigg)- \frac{r}{4 \beta \xi ^2 z^2}\bigg( \pi  \xi  \big(8 \xi ^4 (1-z) z\notag\\&\:-z^2 (2+z)-2 \xi ^2 (4-2 z-2 z^2-z^3)\big)+4 \beta \big(32 \xi ^6 (1-z)-4 \xi ^4 (1-3 z) z\notag\\&\:-\xi ^2 z (4-4 z+3 z^2)+z^3\big)\bigg) \Im^2\bigg[\ln \bigg(1-\frac{z-x^+}{z-x^-}\bigg)\bigg]+ \frac{r}{16 \beta ^2 \xi ^2 z^2} \bigg(48 \xi ^6 (4-5 z\notag\\&\:+z^2)-4 \xi ^4 (80-64 z+4 z^2-3 z^3)+16 \xi ^2 (8-z-2 z^3)-4 z^2 (8-5 z)\notag\\&\:-\beta  \pi \big(24\xi ^5 (1-z) z-6\xi ^3 (4-2 z-2 z^2-z^3)-3 \xi z^2 (2+z)\big)\bigg)\ln ^2(1-z)\notag\\&\:+\frac{1}{8 \xi ^2} \bigg(32 \xi ^4-\xi ^2 (2-21 z)+4 z-\frac{9 \pi  \xi  (2 \xi ^2 z-(2+z))}{\beta }-\frac{128 \xi ^4\beta^2 (1-z) r}{z^2 t}\bigg) \ln ^2(z)\notag\\&\:-\frac{r}{16 \beta ^2 \xi ^2 z^2}\bigg(64 \xi ^6 (-1+z)+8 \beta  \pi  \xi ^5 (-1+z) z+4 (8-5 z) z^2+\beta  \pi  \xi  z^2 (2+z)\notag\\&\:-2 \beta  \pi  \xi ^3 (-4+2 z+2 z^2+z^3)-8 \xi ^4 (-16+6 z+3 z^3)+4 \xi ^2 (-16-4 z-8 z^2\notag\\&\:+11 z^3)\bigg) \ln ^2(1-r)- \frac{r}{2 \beta  \xi ^2 z^2}\bigg(2  \pi  \xi  \big(8 \xi ^4 (1-z) z-2 \xi ^2 (4-2 z-2 z^2-z^3)\notag\\&\:-z^2 (2+z)\big)-\beta \big(16 \xi ^6 (1-z)-4 \xi ^4 (3-6 z+2 z^2)-\xi ^2 z (4-z-z^2)+z^3\big)\bigg) \ln ^2(r)\notag\\&\:+ \frac{r}{4 \xi  z^2} \bigg(9 \pi  \xi  z \big(4 \xi ^2 (1-z)+z^2\big)-2 \beta  \big(32 \xi ^4 (1-z)-4 \xi ^2 (2-3 z-z^2)\notag\\&\:+(2-z) z^2\big)\bigg)\arg \bigg(\frac{x^+}{x^-}\bigg) \ln (1-z)+ \frac{r}{8 \beta  \xi ^2 z^2}\bigg(11 \pi  \xi  \big(8 \xi ^4 (1-z) z-2 \xi ^2 (4-2 z\notag\\&\:-2 z^2-z^3)-z^2 (2+z)\big)-4 \beta  \big(64 \xi ^6 (1-z)-4 \xi ^4 (16-19 z-z^2)+\xi ^2 (32+8 z\notag\\&\:-32 z^2+3 z^3)-4 (1-z) z^2\big)\bigg)\ln (z) \ln (1-z)-\frac{r}{8 \beta ^2 \xi ^2 z^2}\bigg(512 \xi ^8 (-1+z)\notag\\&\:+40 \beta  \pi  \xi ^5 (-1+z) z+5 \beta  \pi  \xi  z^2 (2+z)+4 z^2 (-4+3 z)-32 \xi ^6 (-30+31 z+3 z^2)\notag\\&\:-10 \beta  \pi  \xi ^3 (-4+2 z+2 z^2+z^3)+16 \xi ^4 (-32+33 z+10 z^2+z^3)-4 \xi ^2 (-16+12 z\notag\\&\:+12 z^2+7 z^3)\bigg) \ln (1-r) \ln (1-z)+\frac{r}{4 \beta ^2 \xi ^2 z^2} \bigg(40 \beta  \pi  \xi ^5 (-1+z) z+5 \beta  \pi  \xi  z^2 (2+z)\notag\\&\:+2 z^2 (-4+3 z)+8 \xi ^4 (-8+5 z)+16 \xi ^6 (2-3 z+z^2)+\xi ^2 (32+8 z-8 z^2-6 z^3)\notag\\&\:-10 \beta  \pi  \xi ^3 (-4+2 z+2 z^2+z^3)\bigg) \ln (r) \ln (1-z)-\frac{r}{2 \beta \xi ^2 z^2} \bigg(\pi  \xi  \bigg(8 \xi ^4 (1-z) z\notag\\&\:-2 \xi ^2 (4-2 z-2 z^2-z^3)-z^2 (2+z)\bigg)-2\beta\bigg(64 \xi ^6 (1-z)-4 \xi ^4 (12-13 z-3 z^2)\notag\\&\:-\xi ^2 z (8+8 z-z^2)+(2-z) z^2\bigg)\bigg) \ln (t) \ln (1-z)+\frac{2 r}{\xi  z^2}\bigg(12 \pi  \xi ^3 (1-z) z+3 \pi  \xi  z^3\notag\\&\:+\beta\big(32\xi ^4 (1-z)-4\xi ^2 (2-3 z-z^2)+(2-z) z^2\big)\bigg)\arg \bigg(\frac{x^+}{x^-}\bigg)  \ln (z)\notag\\&\:+\frac{r}{4 \xi  z^2}\bigg(20 \pi  \xi ^3 (1-z) z+5 \pi  \xi  z^3+2\beta\big(32 \xi ^4 (1-z)-4\xi ^2 (2-3 z-z^2)\notag\\&\:+(2-z) z^2\big)\bigg)\arg \bigg(\frac{x^+}{x^-}\bigg) \ln (1-r)- \frac{r}{8 \beta  \xi ^2 z^2} \bigg(7 \pi  \big(8 \xi ^5 (1-z) z-2 \xi ^3 (4-2 z-2 z^2\notag\\&\:-z^3)-\xi  z^2 (2+z)\big)-8 \beta  \big(64 \xi ^6 (1-z)-4 \xi ^4 (12-13 z-3 z^2)-\xi ^2 z (8+8 z-z^2)\notag\\&\:+(2-z) z^2\big)\bigg)\ln (z) \ln (1-r)-\bigg(\frac{5 \pi  z}{2}+\frac{3 \beta r}{2 \xi  z^2}\big(32 \xi ^4 (1-z)-4 \xi ^2 (2-3 z-z^2)\notag\\&\:+(2-z) z^2\big)\bigg)\arg \bigg(\frac{x^+}{x^-}\bigg)  \ln (r)+\frac{r}{2 \beta \xi ^2 z^2}\bigg(4  \pi  \xi  \big(8 \xi ^4 (1-z) z-2 \xi ^2 (4-2 z\notag\\&\:-2 z^2-z^3)-(2+z)z^2\big)-\beta \big(64 \xi ^6 (1-z)-4 \xi ^4 (10-19 z+5 z^2)-\xi ^2 z (8+2 z\notag\\&\:-5 z^2)+2 z^3\big)\bigg) \ln (z) \ln (r)+\frac{r}{4 \beta \xi ^2 z^2}\bigg(3  \pi  \xi  \big(8 \xi ^4 (1-z) z-2 \xi ^2 (4-2 z-2 z^2-z^3)\notag\\&\:-(2+z)z^2\big)-2 \beta \big(64 \xi ^6 (1-z)-16 \xi ^4 (5-5 z-z^2)-2 \xi ^2 z (6+8 z-z^2)\notag\\&\:+z^2 (4-3 z)\big)\bigg) \ln (1-r) \ln (r)+\frac{\pi (2 \xi ^2 z-(2+z))}{4 \beta  \xi } \bigg(\ln (r) \ln \bigg(\bigg(1-\frac{r}{z}\bigg)^2\bigg)\notag\\&\:- \ln (z) \ln \bigg(\bigg(1-\frac{r}{z}\bigg)^2\bigg)\bigg)-\frac{r}{2 \beta ^2 \xi ^2 z^2} \bigg(128 \xi ^8 (1-z)-16 \xi ^6 (12-13 z-z^2)\notag\\&\:+4 \xi ^4 (24-28 z-3 z^3)-4 \xi ^2 (8-8 z+6 z^2-5 z^3)+8 (1-z) z^2-\pi \beta \big(8 \xi ^5 (1-z) z\notag\\&\:-2 \xi ^3 (4-2 z-2 z^2-z^3)- \xi  z^2 (2+z)\big)\bigg) \arg \bigg(\frac{x^+}{x^-}\bigg) \Im\bigg[\ln \bigg(1-\frac{z-x^+}{z-x^-}\bigg)\bigg]\notag\\&\:+ \frac{r}{2 \beta  \xi ^2 z^2} \bigg(\pi  \big(8 \xi ^5 (1-z) z-2 \xi ^3 (4-2 z-2 z^2-z^3)-\xi  z^2 (2+z)\big)-2 \beta  \big(64 \xi ^6 (1-z)\notag\\&\:-4 \xi ^4 (12-13 z-3 z^2)-\xi ^2 z (8+8 z-z^2)+(2-z) z^2\big)\bigg)\bigg(\ln (z) \ln (t)-\ln (r) \ln (t)\bigg)\notag\\&\:-\frac{5\pi  z}{2} \ln (1-z) \Im\bigg[\ln \bigg(1-\frac{z-x^+}{z-x^-}\bigg)\bigg]-7\pi z \ln (z) \Im\bigg[\ln \bigg(1-\frac{z-x^+}{z-x^-}\bigg)\bigg]\notag\\&\:-\frac{3\pi  z}{2} \ln (1-r) \Im\bigg[\ln \bigg(1-\frac{z-x^+}{z-x^-}\bigg)\bigg]+\frac{r}{2 \xi  z^2}\bigg(44 \pi  \xi ^3 (1-z) z+11 \pi  \xi  z^3\notag\\&\:+2\beta\big(32  \xi ^4 (1-z)-4 \xi ^2 (2-3 z-z^2)+(2-z) z^2\big)\bigg) \ln (r) \Im\bigg[\ln \bigg(1-\frac{z-x^+}{z-x^-}\bigg)\bigg]\notag\\&\:+ \frac{r}{24 \beta  \xi ^2 z^2} \bigg(768 \xi ^7 (2-3 z)-8 \xi ^5 \big(216-(372-17 \pi ^2) z-(84+17 \pi ^2) z^2\big)\notag\\&\:+2 \xi ^3 \big(96+68 \pi ^2-(336+34 \pi ^2) z-(408+34 \pi ^2) z^2-(12+17 \pi ^2) z^3\big)\notag\\&\:+\xi  z^2 \big(144+34 \pi ^2+(24+17 \pi ^2) z\big)-24 \beta  \pi  \big(32 \xi ^6 (1-z)-4 \xi ^4 (8-11 z+z^2)\notag\\&\:+\xi ^2 (16-12 z+4 z^2-3 z^3)-(4-3 z) z^2\big)\bigg)\Im\bigg[\ln \bigg(1-\frac{z-x^+}{z-x^-}\bigg)\bigg] \notag\\&\:-\frac{r}{2 \beta  \xi ^2 z^2}\bigg(64 \xi ^7 (2-3 z)-8 \xi ^5 \big(18-(35-\pi ^2) z-(3+\pi ^2) z^2\big)+2 \xi ^3 \big((8+4 \pi ^2)\notag\\&\:-(44+2 \pi ^2) z-(18+2 \pi ^2) z^2-(1+\pi ^2) z^3\big)+\xi  z^2 \big(12+2\pi ^2+(2+\pi^2)z\big)\notag\\&\:-2 \beta  \pi  \big(32 \xi ^6 (1-z)-4 \xi ^4 (8-11 z+z^2)+\xi ^2 (16-4 z-3 z^3)\notag\\&\:-2 (3-2 z) z^2\big)\bigg)\arg \bigg(\frac{x^+}{x^-}\bigg)+\bigg(\frac{\beta  \pi r}{\xi  z^2} \big(32 \xi ^4 (1-z)-4 \xi ^2 (2-3 z-z^2)+(2-z) z^2\big)\notag\\&\:+\frac{z t}{96 \xi ^2 (1-z)}\big(\xi ^2(1-z) (336-116 \pi ^2)-48+(108-29\pi^2)z\big)\bigg) \ln (1-z)\notag\\&\:-\bigg(\frac{5 \pi ^2 z}{8}+\frac{\beta  \pi r}{\xi  z^2}\big(32 \xi ^4 (1-z)-4 \xi ^2 (2-3 z-z^2)+(2-z) z^2\big)\bigg) \ln (1-r)\notag\\&\:+\bigg(\frac{r t}{16 \xi^2 (1-z) z^2 (4 \xi^2-z)^2}\big(8192 \xi^{10} (2-5 z+3 z^2)-256 \xi^8 (5-10 z-83 z^2\notag\\&\:+96 z^3)-64 \xi^6 z (107-110 z+89 z^2-166 z^3)+16 \xi^4  z^2(119-271 z+43 z^2\notag\\&\:-140 z^3)+4 \xi^2 z^3 (125+72 z+53 z^2+46 z^3)-(128-61 z+50 z^2) z^4\big)\notag\\&\:-\frac{\pi ^2}{48 \xi ^2 (1-z) z}\big(400 r t \xi ^4 (1-z)^2+4 \xi ^2 (1-z) z (25 r t -(2-25 r t) z)+25 r t z^3\big)\notag\\&\:-\frac{\pi  \beta r t}{2 \xi ^3 (1-z) z^2}\big(128 \xi ^6 (1-z)^2-16 \xi ^4 (2-7 z+4 z^2+z^3)-4 \xi ^2 z (2-5 z+2 z^2\notag\\&\:-z^3)+(2-z) z^3\big)\bigg) \ln (z)-\bigg(\frac{r t}{8 \xi ^4 (4 \xi ^2-z)^2 (1-z) z^2} \big(2048 \xi ^{12} (2-5 z+3 z^2)\notag\\&\:-256 \xi ^{10} (1+2 z-29 z^2+28 z^3)-64 \xi ^8 z (43-65 z+61 z^2-59 z^3)\notag\\&\:+16 \xi ^6 z^2
(46-109 z+50 z^2-58 z^3)+4 \xi ^4 z^3 (52+5 z+19 z^2+21 z^3)\notag\\&\:-\xi ^2 z^4 (52-32 z+27 z^2)+z^6\big)-\frac{7 \pi ^2 r t}{32 \xi ^2 (1-z) z} \big(16 \xi ^4 (1-2 z+z^2)+4 \xi ^2 z (1-z^2)\notag\\&\:+z^3\big)-\frac{3 \beta  \pi  r t}{8 \xi ^3 (1-z) z^2} \big(128 \xi ^6 (1-2 z+z^2)-16 \xi ^4 (2-7 z+4 z^2+z^3)\notag\\&\:-4 \xi ^2 z (2-5 z+2 z^2-z^3)+(2-z) z^3\big)\bigg)\ln (r)+\bigg(\frac{\pi  r}{16 \beta  \xi  z^2} \big(256 \xi ^6 (2-3 z)\notag\\&\:-8 \xi ^4 (72-(156-3 \pi ^2) z-(-4+3 \pi ^2) z^2)+2 \xi ^2 (32+12 \pi ^2-(240+6 \pi ^2) z\notag\\&\:-(8+6
\pi ^2) z^2-(4+3 \pi ^2) z^3)+z^2 (48+6 \pi ^2+(8+3\pi^2)z)\big)\notag\\&\:-\frac{r}{12 \xi ^2 (4 \xi ^2-z) z^2} \big(384 \xi ^8 \pi^2(1-z)-16 \xi ^6 ((-57+42 \pi ^2)+(174-59 \pi ^2) z\notag\\&\:+(-117+11 \pi ^2) z^2)+4 \xi ^4
(96 \pi ^2+(-243+38 \pi ^2) z+(411-71 \pi ^2) z^2\notag\\&\:+(-276+6 \pi ^2) z^3)-\xi ^2 z (96 \pi ^2+(-192+156 \pi ^2) z+(243-106 \pi ^2) z^2\notag\\&\:+(-159-11
\pi ^2) z^3)-5 \pi ^2 z^3 (-8+5 z)\big)+z \zeta (3)\bigg)\bigg\}\;.
\end{align}

\medskip 

\begin{figure}[t]
	\centering
	\includegraphics[width=0.9\textwidth]{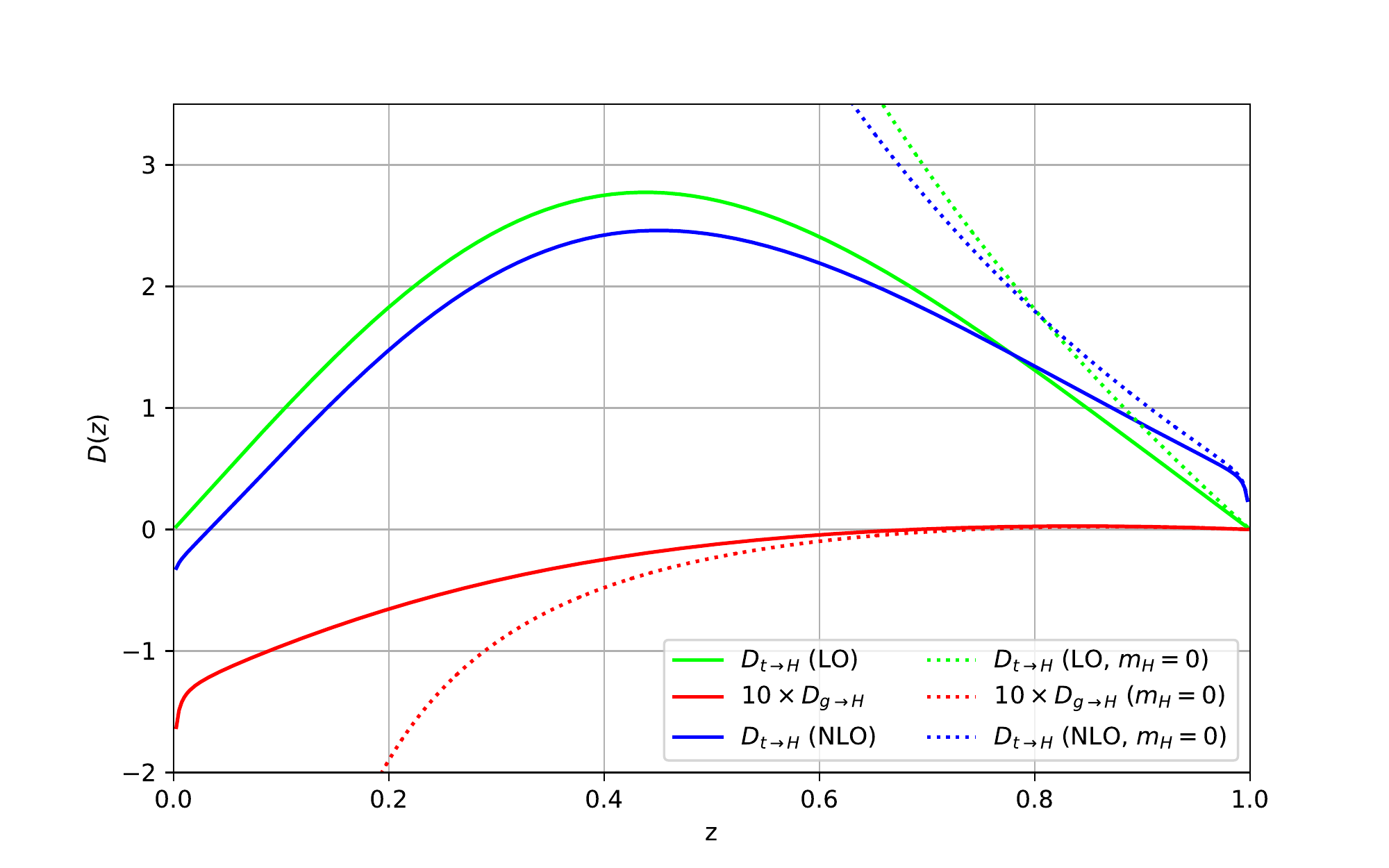}
	\caption{A comparison of the fragmentation functions $D_{g\to H}$ (red) and $D_{t\to H}$ (blue) at NLO with $D_{t\to H}$ (green) at LO. The solid lines correspond to $m_H = 125$ GeV, while the dotted lines show the results for $m_H = 0$. In both cases, $\mu = m_t = 173$ GeV and $\alpha_s(m_t) = 0.108$, while the overall factor $y_t^2/(16\pi^2)$ is set to $1$. Note that $D_{g\to H}$ has been magnified by a factor of 10.}
	\label{fig:GluVsLO}
\end{figure}

Fig.\ \ref{fig:GluVsLO} shows a comparison between the NLO results and the LO top-to-Higgs FF. The NLO corrections to $D_{t\to H}$ are roughly $\mathcal{O}(10\%)$ compared to the LO result. For both $z\to0$ and $z\to1$, the NLO corrections are logarithmically divergent (for $m_H>0$). However, the coefficients of the divergences are numerically tiny, so that they only become relevant for values of $z$ extremely close to $0$ or $1$, as is apparent from fig.\ \ref{fig:GluVsLO}.

The fragmentation function $D_{g\to H}$ is also logarithmically divergent for $z\to0$, while for $z\to1$ it tends to $0$. As the magnitude of the gluon-to-Higgs FF is consistently much smaller than the top-to-Higgs FF, gluon-induced contributions can safely be neglected as long as their suppression is not compensated for by an enhanced partonic cross section, in analogy to the non-singlet approximation commonly used for heavy-quark fragmentation. While for hadron collisions the gluon-induced cross section is indeed enhanced with respect to the top-quark production cross section, an approximation where gluon-induced contributions are neglected is appropriate for lepton collisions.

One might be tempted to compute higher-order corrections to the Higgs FFs in the $m_H=0$ limit as a first approximation. While $m_H\ll m_t$ is not satisfied in nature, this has nonetheless proven to be a surprisingly good approximation in other calculations, see e.g.\ Ref.\ \cite{deFlorian:2016spz} for a review. However, the LO result already demonstrates that this is not the case here. Indeed, the form of the LO FF suggests that $m_H/(z m_t)$ would be a more appropriate expansion variable. The same behaviour holds at NLO, suggesting that the $m_H=0$ limit is only a good approximation for large $z\gtrsim0.7$. This is born out by the numerical results for the NLO fragmentation functions corresponding to $m_H=0$ and $m_H>0$ shown in fig.\ \ref{fig:GluVsLO}.

\section{Conclusions}\label{sec:Conclusions}
In this paper we have presented the fragmentation and splitting functions for the final-state transition from a top-quark to a Higgs boson or from a gluon to a Higgs boson through NLO in QCD, i.e.\ through $\mathcal{O}(y_t^2\alpha_s)$. These new results can be used in two different ways to obtain higher-precision theoretical predictions for any process involving these transitions. First, any such process can be described at NLO in QCD, up to power corrections in the top-quark or Higgs-boson mass, by performing the convolution of massless coefficient functions with the fragmentation functions. As setting all masses to zero can significantly simplify the computation of coefficient functions, this may allow to obtain higher-order results which would be prohibitively complicated for the massive case. Second, the results can be used to perform a resummation of logarithms of the type $\ln(p_T/m)$ at NLL accuracy using the DGLAP equations, which is necessary to maintain the perturbative convergence of the predictions for top-Higgs associated production at large transverse momentum.

To obtain our results, we started from the field-theoretic definitions of the fragmentation functions given in Ref.\ \cite{Collins:1981uw}. All employed computational techniques (e.g.\ reverse unitarity, IBP reduction and the solution of master integrals using differential equations) have become increasingly standard for multi-loop calculations in recent years, and most of them were already applied in Refs.\ \cite{Melnikov:2004bm,Mitov:2004du} to the computation of the NNLO corrections to the heavy-quark fragmentation functions.

It has been demonstrated that in order to obtain accurate results, one cannot neglect the Higgs-boson mass in the computation of the Higgs fragmentation functions. At the same time, the results suggest that neglecting fragmentation contributions from gluons may be justified at lepton-colliders. In subsequent work we plan to study the size of missing power corrections, the effect of NLL DGLAP resummation and the size of gluon-initiated contributions for top-Higgs associated production at the LHC, to explore the phenomenological importance of those effects.

\begin{acknowledgments}
We would like to thank Alexander M\"uck and Michael Spira for discussions. The work of C.B., M.C.\ and M.K.\ was supported by the Deutsche Forschungsgemeinschaft (DFG) under grant 396021762 - TRR 257: Particle Physics Phenomenology after the Higgs Discovery. The work of T.G.\ was supported by the DFG under grant 400140256 - GRK 2497: The physics of the heaviest particles at the Large Hadron Collider.
\end{acknowledgments}

\clearpage

\appendix

\renewcommand{\thefigure}{A\arabic{figure}}
\setcounter{figure}{0}

\section{Feynman diagrams}\label{sec:FeynDia}

\begin{figure}[h]
	\centering
	\begin{subfigure}{0.31\textwidth}
		\includegraphics[width=0.9\textwidth]{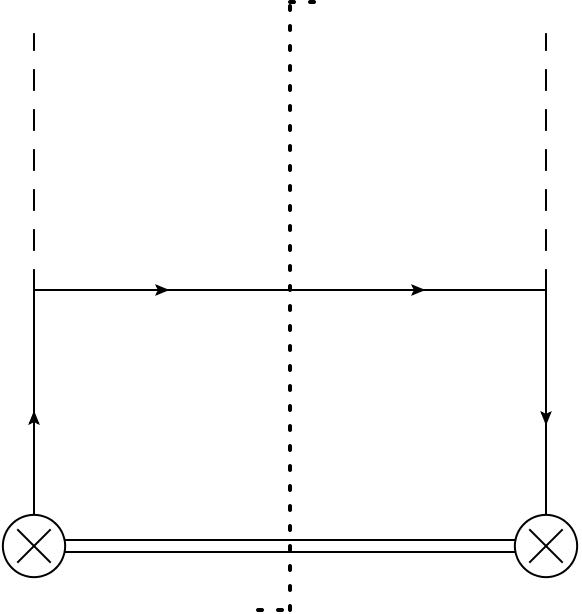}
		\caption{}
	\end{subfigure}
	\begin{subfigure}{0.31\textwidth}
		\includegraphics[width=0.9\textwidth]{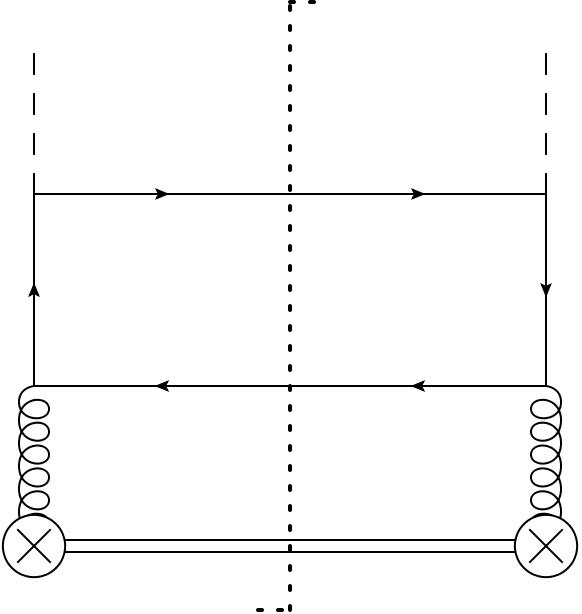}
		\caption{}
	\end{subfigure}
	\begin{subfigure}{0.31\textwidth}
		\includegraphics[width=0.9\textwidth]{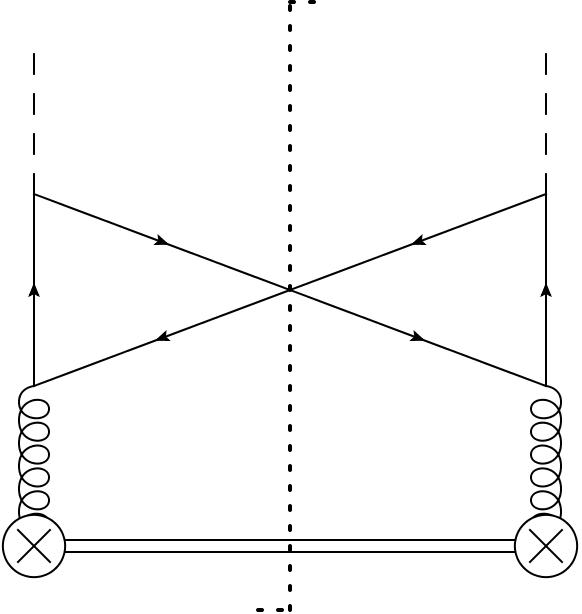}
		\caption{}
	\end{subfigure}
	\caption{The LO diagram for $D_{t\to H}$ (a) and the two diagrams for $D_{g\to H}$ at NLO ((b) and (c)). There are also conjugate diagrams to (b) and (c). Lines with arrows denote (anti-)top quarks, curly lines denote gluons and dashed lines denote Higgs bosons. The dotted line in the centre shows the cut, while the double line at the bottom represents the Wilson line. The Higgs-boson line does not cross the cut to symbolise that its momentum is not integrated over. \cite{JaxoDraw}}
	\label{fig:LOandGlu}
\end{figure}
\begin{figure}[h]
	\centering
	\begin{subfigure}{0.31\textwidth}
		\includegraphics[width=0.9\textwidth]{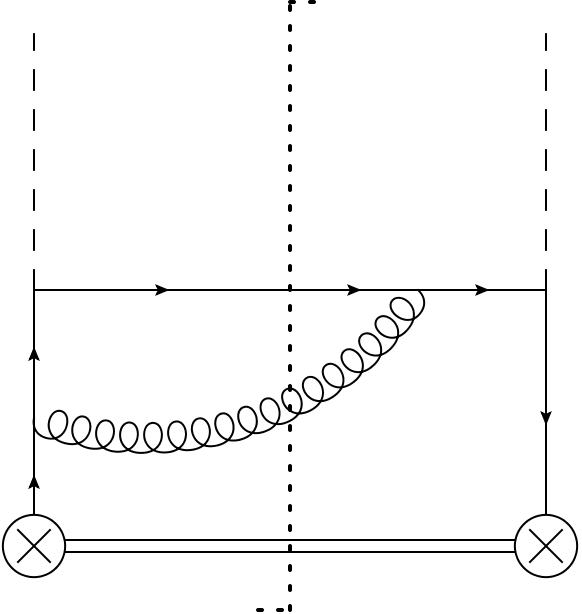}
		\caption{}
	\end{subfigure}
	\begin{subfigure}{0.31\textwidth}
		\includegraphics[width=0.9\textwidth]{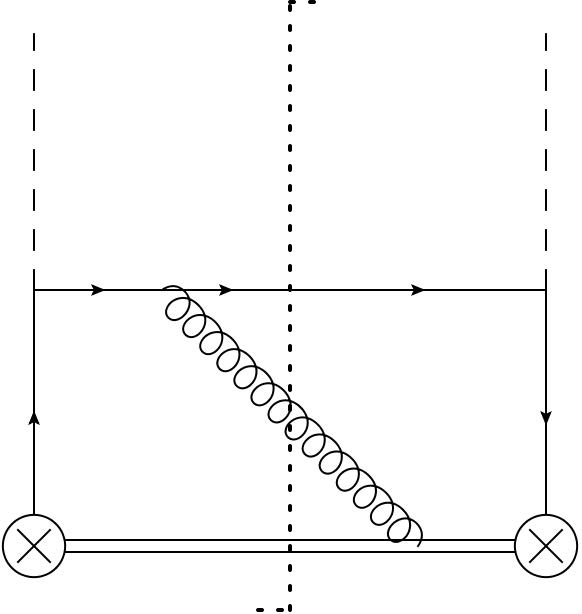}
		\caption{}
	\end{subfigure}
	\begin{subfigure}{0.31\textwidth}
		\includegraphics[width=0.9\textwidth]{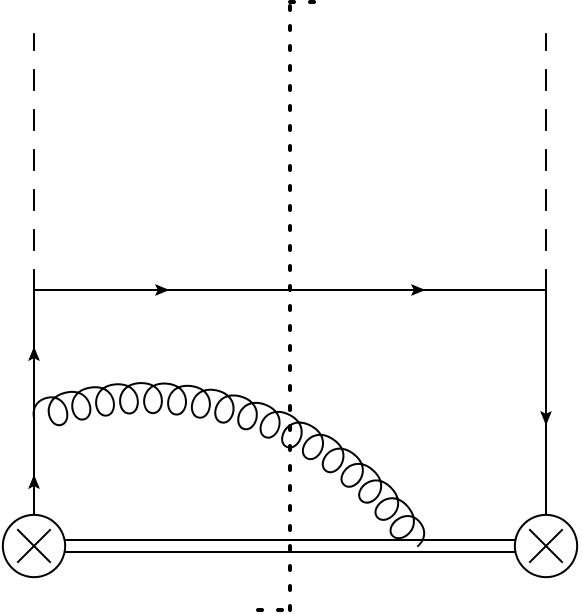}
		\caption{}
	\end{subfigure}
	\begin{subfigure}{0.31\textwidth}
		\includegraphics[width=0.9\textwidth]{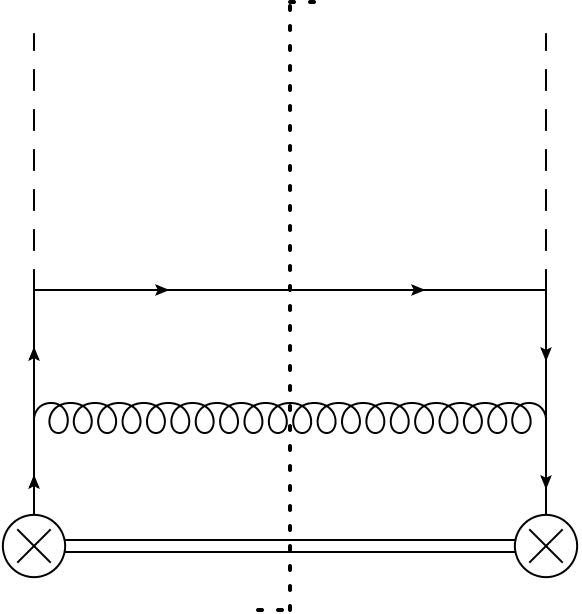}
		\caption{}
	\end{subfigure}
	\begin{subfigure}{0.31\textwidth}
		\includegraphics[width=0.9\textwidth]{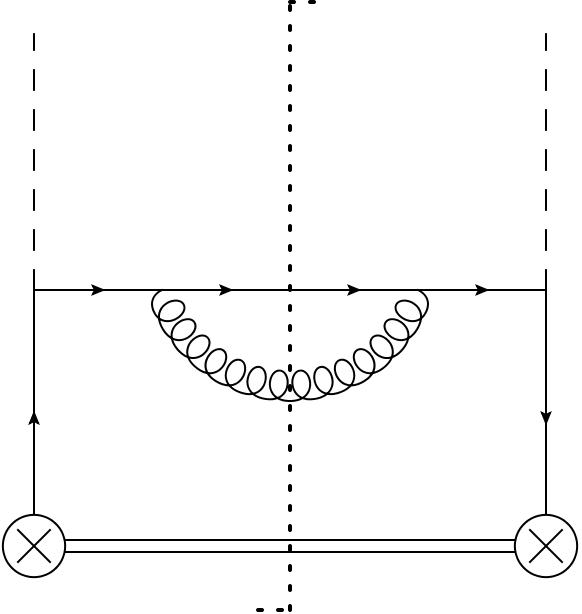}
		\caption{}
	\end{subfigure}
	\caption{The diagrams for the real corrections to $D_{t\to H}$. There are also conjugate diagrams to (a), (b) and (c).}
	\label{fig:TopReal}
\end{figure}

\begin{figure}[b]
	\centering
	\begin{subfigure}{0.31\textwidth}
		\includegraphics[width=0.9\textwidth]{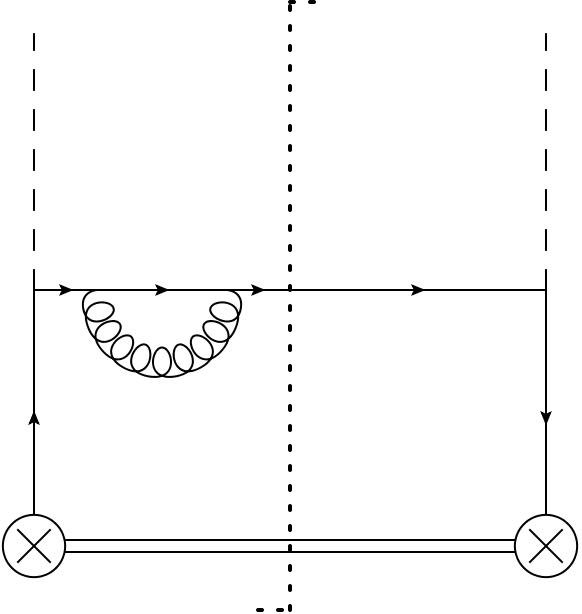}
		\caption{}
	\end{subfigure}
	\begin{subfigure}{0.31\textwidth}
		\includegraphics[width=0.9\textwidth]{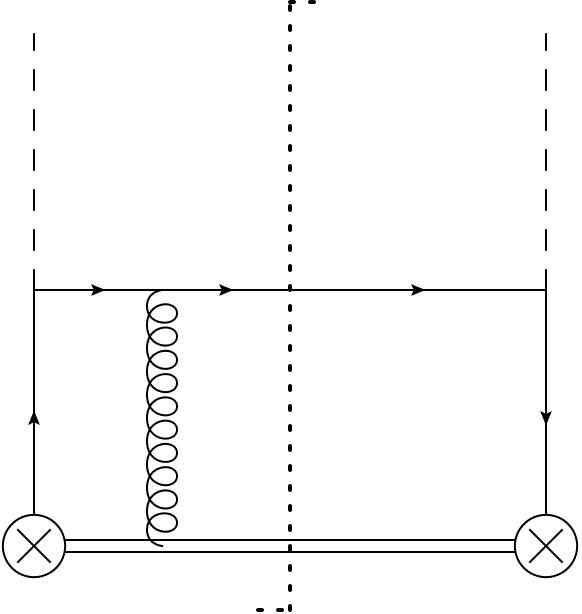}
		\caption{}
	\end{subfigure}
	\begin{subfigure}{0.31\textwidth}
		\includegraphics[width=0.9\textwidth]{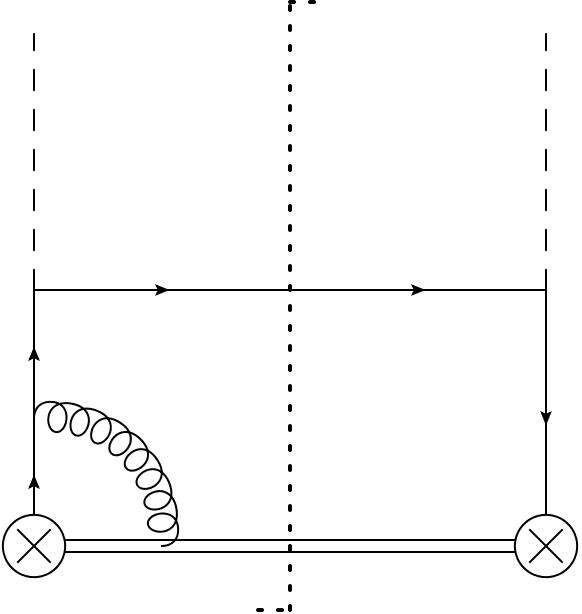}
		\caption{}
	\end{subfigure}
	\begin{subfigure}{0.31\textwidth}
		\includegraphics[width=0.9\textwidth]{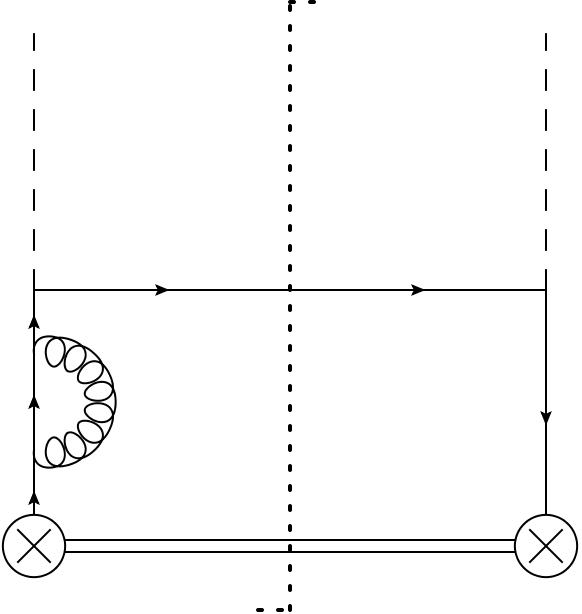}
		\caption{}
	\end{subfigure}
	\begin{subfigure}{0.31\textwidth}
		\includegraphics[width=0.9\textwidth]{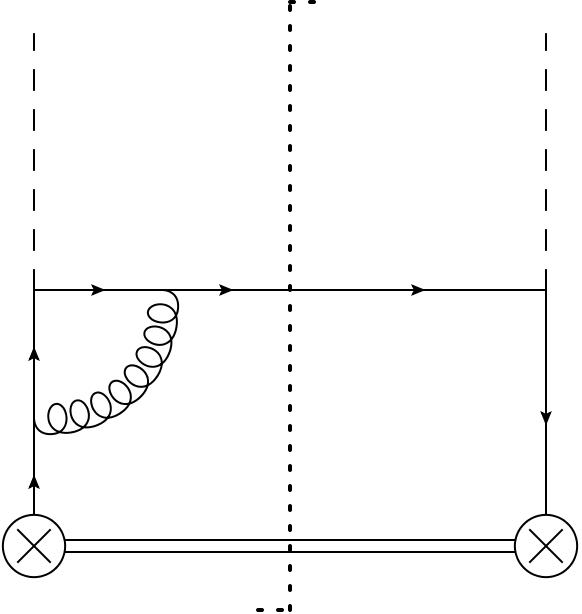}
		\caption{}
	\end{subfigure}
	\caption{The diagrams for the virtual corrections to $D_{t\to H}$. All diagrams have a conjugate diagram which is not shown.}
	\label{fig:TopVirt}
\end{figure}

\clearpage

\section{Feynman diagrams for master integrals}\label{sec:FeynDiaMIs}

\begin{figure}[h!]
    \centering
    \includegraphics[]{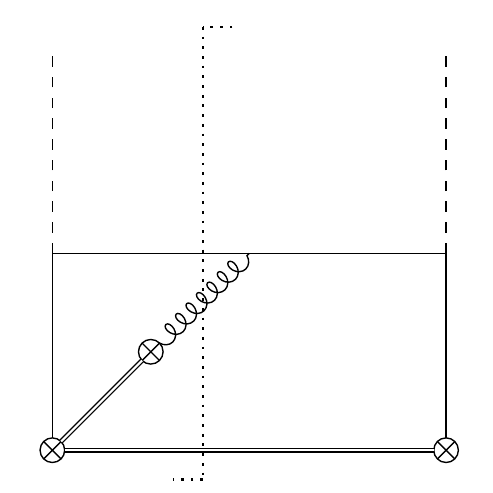}
    \caption{The scalar integral topology for the real corrections to $D_{t \to H}$ defined in eq.\ \eqref{eq:TopoReal}. Solid lines are massive and correspond to top-quark propagators, the curly line corresponds to the gluon, dashed lines to the Higgs boson and double lines are Wilson lines. The dotted line shows the cut.}
    \label{fig:RealTopo}
\end{figure}

\begin{figure}[h!]
    \centering
    \includegraphics[]{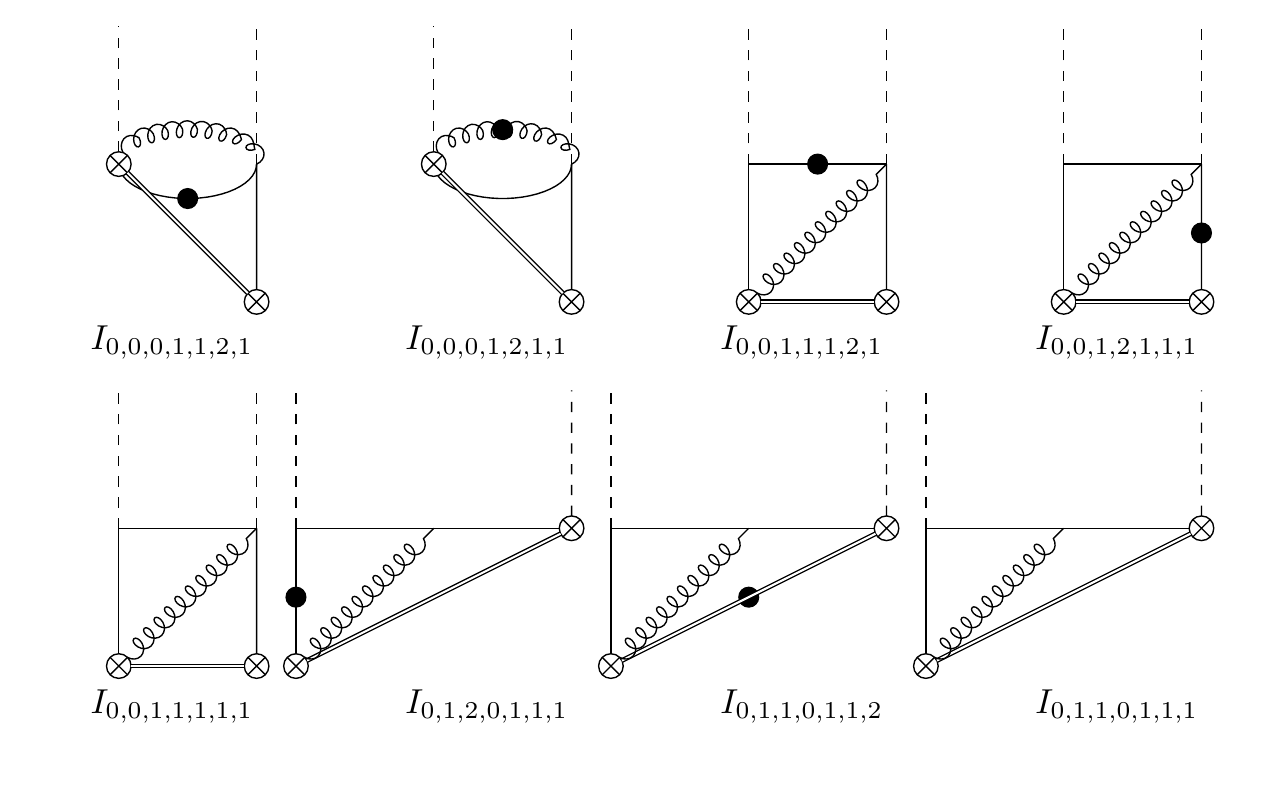}
    \caption{The pre-canonical master integrals  for the real corrections to $D_{t \to H}$. The black dots represent squared propagators.}
    \label{fig:RealMIs}
\end{figure}

\begin{figure}[h!]
    \centering
    \includegraphics[]{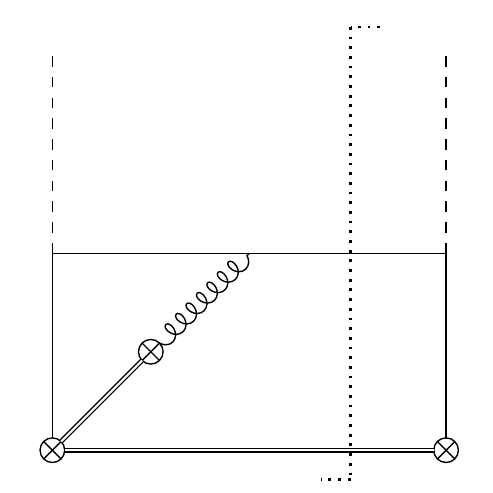}
    \caption{The scalar integral topology for the virtual corrections to $D_{t \to H}$ defined in eq.\ \eqref{eq:TopoVirt}.}
    \label{fig:VirtTopo}
\end{figure}

\begin{figure}[h!]
    \centering
    \includegraphics[]{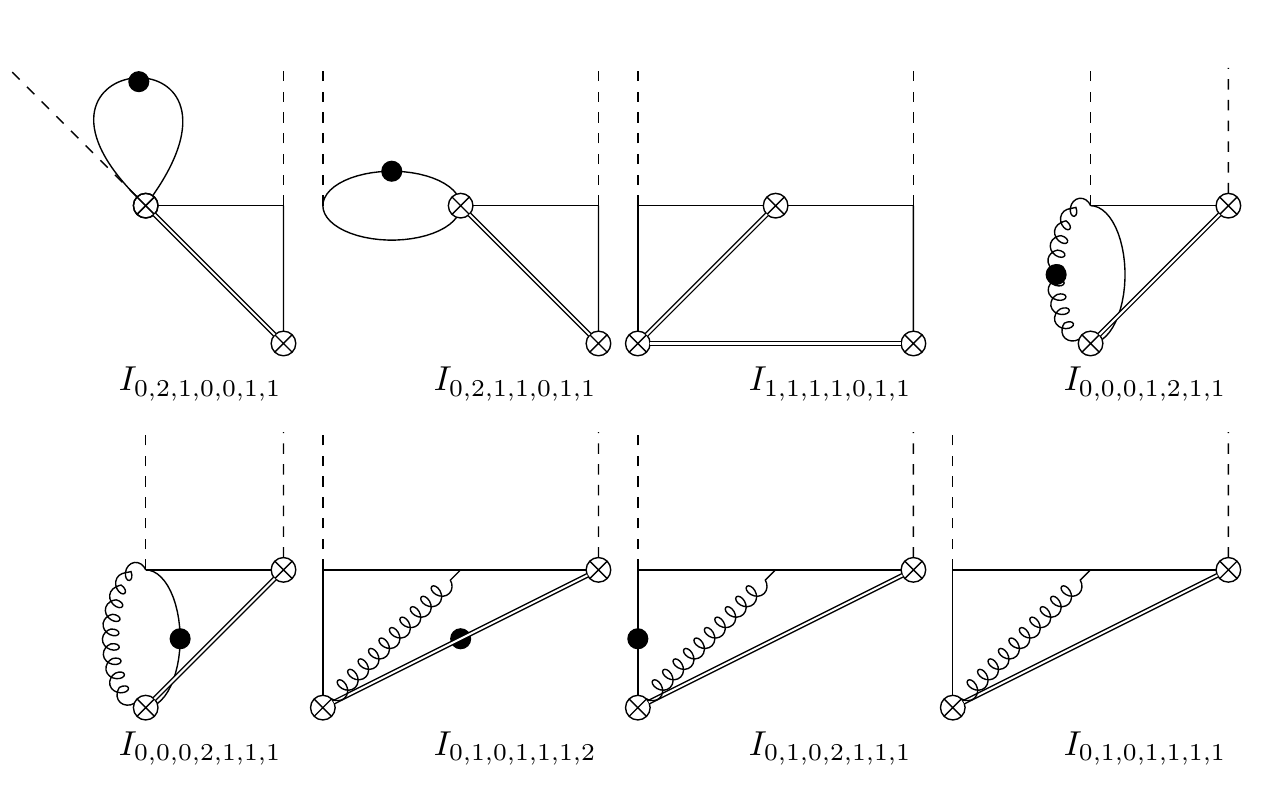}
    \caption{The pre-canonical master integrals  for the virtual corrections to $D_{t \to H}$.}
    \label{fig:VirtMIs}
\end{figure}

\clearpage

\section{Canonical form}\label{sec:CanMat}

\subsection{Canonical form for real corrections}
The differential equation matrix in $\mathrm{d}\!\log$ form for the real corrections is:
\begin{equation}
\begin{aligned}
\mathrm{d}\Tilde{A}({\tau}) & = M_1 \, \mathrm{d}\!\log\left(\tau\right)
+ M_2 \, \mathrm{d}\!\log\left(1-\tau\right)
+ M_3 \, \mathrm{d}\!\log\left(1 + \tau\right) \\
& + M_4 \, \mathrm{d}\!\log\left(2-\left(1-\tau \right) z\right)
+ M_5 \, \mathrm{d}\!\log\left(-2+\left(1+\tau\right) z\right) \\
&+ M_6 \, \mathrm{d}\!\log\left(4-\left(1-\tau ^2\right) z \right)\;,
\end{aligned}
\end{equation}
with
\begin{equation}
M_{1}=\left(
\begin{array}{cccccccc}
 0 & 0 & 0 & 0 & 0 & 0 & 0 & 0 \\
 0 & 0 & 0 & 0 & 0 & 0 & 0 & 0 \\
 0 & 0 & -2 & 0 & 0 & 0 & 0 & 0 \\
 0 & 0 & 0 & 0 & 0 & 0 & 0 & 0 \\
 0 & 0 & 0 & 0 & 0 & 0 & 0 & 0 \\
 0 & 0 & 0 & 0 & 0 & -2 & 0 & 0 \\
 0 & 0 & 0 & 0 & 0 & 0 & 0 & 0 \\
 0 & 0 & 0 & 0 & 0 & 0 & 0 & 0 \\
\end{array}
\right)\;,
\end{equation}
\begin{equation}
M_{2}=\left(
\begin{array}{cccccccc}
 -1 & 1 & 0 & 0 & 0 & 0 & 0 & 0 \\
 0 & 0 & 0 & 0 & 0 & 0 & 0 & 0 \\
 i & -i & 0 & -2 i & 0 & 0 & 0 & 0 \\
 0 & 0 & 0 & -1 & 0 & 0 & 0 & 0 \\
 -\frac{1}{2} & -\frac{1}{2} & \frac{i}{2} & 1 & -1 & 0 & 0 & 0 \\
 0 & 0 & 0 & 0 & 0 & 0 & i & 2 i \\
 0 & 0 & 0 & 0 & 0 & -i & 0 & 2 \\
 0 & 0 & 0 & 0 & 0 & \frac{i}{2} & -\frac{1}{2} & -2 \\
\end{array}
\right)\;,
\end{equation}
\begin{equation}
M_{3}=\left(
\begin{array}{cccccccc}
 -1 & 1 & 0 & 0 & 0 & 0 & 0 & 0 \\
 0 & 0 & 0 & 0 & 0 & 0 & 0 & 0 \\
 -i & i & 0 & 2 i & 0 & 0 & 0 & 0 \\
 0 & 0 & 0 & -1 & 0 & 0 & 0 & 0 \\
 -\frac{1}{2} & -\frac{1}{2} & -\frac{i}{2} & 1 & -1 & 0 & 0 & 0 \\
 0 & 0 & 0 & 0 & 0 & 0 & -i & -2 i \\
 0 & 0 & 0 & 0 & 0 & i & 0 & 2 \\
 0 & 0 & 0 & 0 & 0 & -\frac{i}{2} & -\frac{1}{2} & -2 \\
\end{array}
\right)\;,
\end{equation}
\begin{equation}
M_{4}=\left(
\begin{array}{cccccccc}
 0 & 0 & 0 & 0 & 0 & 0 & 0 & 0 \\
 -1 & -3 & 0 & 0 & 0 & 0 & 0 & 0 \\
 0 & -2 i & -1 & 4 i & 2 i & 0 & 0 & 0 \\
 -\frac{1}{2} & -\frac{1}{2} & -\frac{i}{2} & -2 & -1 & 0 & 0 & 0 \\
 0 & 0 & 0 & 0 & 0 & 0 & 0 & 0 \\
 0 & 0 & 0 & 0 & 0 & -1 & 2 i & 2 i \\
 0 & 0 & 0 & 0 & 0 & -i & -2 & -2 \\
 0 & 0 & 0 & 0 & 0 & 0 & 0 & 0 \\
\end{array}
\right)\;,
\end{equation}
\begin{equation}
M_{5}=\left(
\begin{array}{cccccccc}
 0 & 0 & 0 & 0 & 0 & 0 & 0 & 0 \\
 -1 & -3 & 0 & 0 & 0 & 0 & 0 & 0 \\
 0 & 2 i & -1 & -4 i & -2 i & 0 & 0 & 0 \\
 -\frac{1}{2} & -\frac{1}{2} & \frac{i}{2} & -2 & -1 & 0 & 0 & 0 \\
 0 & 0 & 0 & 0 & 0 & 0 & 0 & 0 \\
 0 & 0 & 0 & 0 & 0 & -1 & -2 i & -2 i \\
 0 & 0 & 0 & 0 & 0 & i & -2 & -2 \\
 0 & 0 & 0 & 0 & 0 & 0 & 0 & 0 \\
\end{array}
\right)\;,
\end{equation}
\begin{equation}
M_{6}=\left(
\begin{array}{cccccccc}
 -1 & -1 & 0 & 0 & 0 & 0 & 0 & 0 \\
 1 & 1 & 0 & 0 & 0 & 0 & 0 & 0 \\
 0 & 0 & 0 & 0 & 0 & 0 & 0 & 0 \\
 1 & 1 & 0 & 0 & 0 & 0 & 0 & 0 \\
 0 & 0 & 0 & 0 & 0 & 0 & 0 & 0 \\
 0 & 0 & 0 & 0 & 0 & 0 & 0 & 0 \\
 0 & 0 & 0 & 0 & 0 & 0 & 0 & 0 \\
 0 & 0 & 0 & 0 & 0 & 0 & 0 & 0 \\
\end{array}
\right)\;.
\end{equation}

\subsection{Canonical form for virtual corrections}
The differential equation matrix in $\mathrm{d}\!\log$ form for the virtual corrections is:
\begin{equation}
\begin{aligned}
\mathrm{d}\Tilde{A}({\tau}) & = M_1 \, \mathrm{d}\!\log\left(\tau\right) 
+ M_2 \, \mathrm{d}\!\log\left(1-\tau\right)
+ M_3 \, \mathrm{d}\!\log\left(1 + \tau\right) \\
& + M_4 \, \mathrm{d}\!\log\left(2-\left(1-\tau \right) z\right) + M_5 \, \mathrm{d}\!\log\left(-2+\left(1+\tau\right) z\right) \\
& + M_6 \, \mathrm{d}\!\log\left(-4+\left(3+\tau ^2\right) z\right)\;,
\end{aligned}
\end{equation}
with
\begin{equation}
M_{1}=\left(
\begin{array}{cccccccc}
 0 & 0 & 0 & 0 & 0 & 0 & 0 & 0 \\
 0 & -2 & 0 & 0 & 0 & 0 & 0 & 0 \\
 0 & 0 & 0 & 0 & 0 & 0 & 0 & 0 \\
 0 & 0 & 0 & 0 & 0 & 0 & 0 & 0 \\
 0 & 0 & 0 & 0 & 0 & 0 & 0 & 0 \\
 0 & 0 & 0 & 0 & 0 & 0 & 0 & 0 \\
 0 & 0 & 0 & 0 & 0 & 0 & -2 & 0 \\
 0 & 0 & 0 & 0 & 0 & 0 & 0 & 0 \\
\end{array}
\right)\;,
\end{equation}
\begin{equation}
M_{2}=\left(
\begin{array}{cccccccc}
 -1 & 0 & 0 & 0 & 0 & 0 & 0 & 0 \\
 i & 0 & 0 & 0 & 0 & 0 & 0 & 0 \\
 0 & 0 & 0 & 0 & 0 & 0 & 0 & 0 \\
 0 & 0 & 0 & 0 & 0 & 0 & 0 & 0 \\
 0 & 0 & 0 & 1 & -1 & 0 & 0 & 0 \\
 0 & 0 & 0 & 1 & 1 & 0 & -i & 2 \\
 0 & 0 & 0 & -2 i & i & i & 0 & 2 i \\
 0 & 0 & 0 & 0 & -\frac{1}{2} & -\frac{1}{2} & \frac{i}{2} & -2 \\
\end{array}
\right)\;,
\end{equation}
\begin{equation}
M_{3}=\left(
\begin{array}{cccccccc}
 -1 & 0 & 0 & 0 & 0 & 0 & 0 & 0 \\
 -i & 0 & 0 & 0 & 0 & 0 & 0 & 0 \\
 0 & 0 & 0 & 0 & 0 & 0 & 0 & 0 \\
 0 & 0 & 0 & 0 & 0 & 0 & 0 & 0 \\
 0 & 0 & 0 & 1 & -1 & 0 & 0 & 0 \\
 0 & 0 & 0 & 1 & 1 & 0 & i & 2 \\
 0 & 0 & 0 & 2 i & -i & -i & 0 & -2 i \\
 0 & 0 & 0 & 0 & -\frac{1}{2} & -\frac{1}{2} & -\frac{i}{2} & -2 \\
\end{array}
\right)\;,
\end{equation}
\begin{equation}
M_{4}=\left(
\begin{array}{cccccccc}
 -1 & 0 & 0 & 0 & 0 & 0 & 0 & 0 \\
 0 & -1 & 0 & 0 & 0 & 0 & 0 & 0 \\
 0 & -i & -2 & 0 & 0 & 0 & 0 & 0 \\
 0 & 0 & 0 & -3 & -1 & 0 & 0 & 0 \\
 0 & 0 & 0 & 0 & 0 & 0 & 0 & 0 \\
 0 & 0 & 0 & 0 & 0 & -2 & -i & -2 \\
 0 & 0 & 0 & 0 & 0 & 2 i & -1 & 2 i \\
 0 & 0 & 0 & 0 & 0 & 0 & 0 & 0 \\
\end{array}
\right)\;,
\end{equation}
\begin{equation}
M_{5}=\left(
\begin{array}{cccccccc}
 -1 & 0 & 0 & 0 & 0 & 0 & 0 & 0 \\
 0 & -1 & 0 & 0 & 0 & 0 & 0 & 0 \\
 0 & i & -2 & 0 & 0 & 0 & 0 & 0 \\
 0 & 0 & 0 & -3 & -1 & 0 & 0 & 0 \\
 0 & 0 & 0 & 0 & 0 & 0 & 0 & 0 \\
 0 & 0 & 0 & 0 & 0 & -2 & i & -2 \\
 0 & 0 & 0 & 0 & 0 & -2 i & -1 & -2 i \\
 0 & 0 & 0 & 0 & 0 & 0 & 0 & 0 \\
\end{array}
\right)\;,
\end{equation}
\begin{equation}
M_{6}=\left(
\begin{array}{cccccccc}
 0 & 0 & 0 & 0 & 0 & 0 & 0 & 0 \\
 0 & 0 & 0 & 0 & 0 & 0 & 0 & 0 \\
 0 & 0 & 0 & 0 & 0 & 0 & 0 & 0 \\
 0 & 0 & 0 & 1 & 1 & 0 & 0 & 0 \\
 0 & 0 & 0 & -1 & -1 & 0 & 0 & 0 \\
 0 & 0 & 0 & -1 & -1 & 0 & 0 & 0 \\
 0 & 0 & 0 & 0 & 0 & 0 & 0 & 0 \\
 0 & 0 & 0 & 0 & 0 & 0 & 0 & 0 \\
\end{array}
\right)\;.
\end{equation}

\clearpage

\section{Feynman Integrals}\label{sec:FeynInts}
\subsection{Masters for the real corrections}
The master integrals for the real corrections are defined as follows:
\begin{align}
I_{a,b,c,d,e,f,g}&\:=-16\pi i\overline{\mu}^{4\epsilon}\int \frac{d^dp_t}{(2\pi)^d}\frac{d^dp_g}{(2\pi)^d}\frac{(n\cdot p_H)^{a+g}}{(n\cdot p_g)^a((p_t+p_g)^2-m_t^2)^b((p_t+p_H)^2-m_t^2)^c}\notag\\&\:\times\frac{1}{((p_t+p_g+p_H)^2-m_t^2)^d(p_g^2)_c^e(p_t^2-m_t^2)_c^f(n\cdot(p_t+p_g+p_H)-z^{-1}n\cdot p_H)_c^g}\;,
\label{eq:TopoReal}
\end{align}
where the prefactor $-16\pi i\overline{\mu}^{4\epsilon}$ has been chosen to match the factor resulting from the use of reverse unitary and the couplings in the full result. Note that the numerator is merely a constant introduced to eliminate overall factors of $n\cdot p_H$. A propagator with a subscript $c$ denotes a cut propagator:
\begin{equation}
\frac{1}{(x)_c^p}\equiv\frac{1}{(x+i\varepsilon)^p}-\frac{1}{(x-i\varepsilon)^p}\;.
\end{equation}
In this notation, the masters are:
\begin{align}
&I_{0,0,1,1,1,1,1}\;,\;\;\;\;I_{0,0,1,2,1,1,1}\;,\;\;\;\;I_{0,0,1,1,1,2,1}\;,\;\;\;\;I_{0,0,0,1,1,2,1}\;,\notag\\&I_{0,0,0,1,2,1,1}\;,\;\;\;\;I_{0,1,1,0,1,1,1}\;,\;\;\;\;I_{0,1,2,0,1,1,1}\;,\;\;\;\;I_{0,1,1,0,1,1,2}\;.
\end{align}
Their expansions in $\epsilon$ through the order required for the present calculation are given below. They are also included as ancillary files with the e-print version of this paper.


\subsection{Masters for the virtual corrections}
The master integrals for the virtual corrections are defined as follows:
\begin{align}
I_{a,b,c,d,e,f,g}&\:=-32\pi^2i\overline{\mu}^{4\epsilon}\int \frac{d^dp_t}{(2\pi)^d}\frac{d^dp_g}{(2\pi)^d}\frac{(n\cdot p_H)^{a+g}}{(n\cdot p_g)^a((p_t-p_g)^2-m_t^2)^b((p_t+p_H)^2-m_t^2)^c}\notag\\&\:\times\frac{1}{((p_t+p_H-p_g)^2-m_t^2)^d(p_g^2)^e(p_t^2-m_t^2)_c^f(n\cdot(p_t+p_H)-z^{-1}n\cdot p_H)_c^g}\;,
\label{eq:TopoVirt}
\end{align}
where the prefactor $-32\pi^2i\overline{\mu}^{4\epsilon}$ has again been chosen to match the factor resulting from reverse unitary and the couplings in the full result. As for the masters for the real corrections, the numerator is merely a constant introduced to eliminate overall factors of $n\cdot p_H$. In this notation, the masters are:
\begin{align}
&I_{0,1,0,1,1,1,1}\;,\;\;\;\;I_{0,1,0,2,1,1,1}\;,\;\;\;\;I_{0,1,0,1,1,1,2}\;,\;\;\;\;I_{1,1,1,1,0,1,1}\;,\notag\\&I_{0,2,1,1,0,1,1}\;,\;\;\;\;I_{0,2,1,0,0,1,1}\;,\;\;\;\;I_{0,0,0,2,1,1,1}\;,\;\;\;\;I_{0,0,0,1,2,1,1}\;.
\end{align}
Their expansions in $\epsilon$ through the order required for the present calculation are given below. They are also included as ancillary files with the e-print version of this paper.



\begin{thebibliography}{99}

\bibitem{Sirunyan:2020eds}
A.~M.~Sirunyan \textit{et al.} [CMS],
Phys. Rev. D \textbf{102}, no.9, 092013 (2020)
doi:10.1103/PhysRevD.102.092013
[arXiv:2009.07123 [hep-ex]].

\bibitem{Sirunyan:2018koj}
A.~M.~Sirunyan \textit{et al.} [CMS],
Eur. Phys. J. C \textbf{79}, no.5, 421 (2019)
doi:10.1140/epjc/s10052-019-6909-y
[arXiv:1809.10733 [hep-ex]].

\bibitem{Aad:2015gba}
G.~Aad \textit{et al.} [ATLAS],
Eur. Phys. J. C \textbf{76}, no.1, 6 (2016)
doi:10.1140/epjc/s10052-015-3769-y
[arXiv:1507.04548 [hep-ex]].

\bibitem{Sirunyan:2018hoz}
A.~M.~Sirunyan \textit{et al.} [CMS],
Phys. Rev. Lett. \textbf{120}, no.23, 231801 (2018)
doi:10.1103/PhysRevLett.120.231801
[arXiv:1804.02610 [hep-ex]].

\bibitem{Aaboud:2018urx}
M.~Aaboud \textit{et al.} [ATLAS],
Phys. Lett. B \textbf{784}, 173-191 (2018)
doi:10.1016/j.physletb.2018.07.035
[arXiv:1806.00425 [hep-ex]].

\bibitem{Beenakker:2001rj}
W.~Beenakker, S.~Dittmaier, M.~Kramer, B.~Plumper, M.~Spira and P.~M.~Zerwas,
Phys. Rev. Lett. \textbf{87}, 201805 (2001)
doi:10.1103/PhysRevLett.87.201805
[arXiv:hep-ph/0107081 [hep-ph]].

\bibitem{Beenakker:2002nc}
W.~Beenakker, S.~Dittmaier, M.~Kramer, B.~Plumper, M.~Spira and P.~M.~Zerwas,
Nucl. Phys. B \textbf{653}, 151-203 (2003)
doi:10.1016/S0550-3213(03)00044-0
[arXiv:hep-ph/0211352 [hep-ph]].

\bibitem{Reina:2001bc}
L.~Reina, S.~Dawson and D.~Wackeroth,
Phys. Rev. D \textbf{65}, 053017 (2002)
doi:10.1103/PhysRevD.65.053017
[arXiv:hep-ph/0109066 [hep-ph]].

\bibitem{Reina:2001sf}
L.~Reina and S.~Dawson,
Phys. Rev. Lett. \textbf{87}, 201804 (2001)
doi:10.1103/PhysRevLett.87.201804
[arXiv:hep-ph/0107101 [hep-ph]].

\bibitem{Dawson:2002tg}
S.~Dawson, L.~H.~Orr, L.~Reina and D.~Wackeroth,
Phys. Rev. D \textbf{67}, 071503 (2003)
doi:10.1103/PhysRevD.67.071503
[arXiv:hep-ph/0211438 [hep-ph]].

\bibitem{Dawson:2003zu}
S.~Dawson, C.~Jackson, L.~H.~Orr, L.~Reina and D.~Wackeroth,
Phys. Rev. D \textbf{68}, 034022 (2003)
doi:10.1103/PhysRevD.68.034022
[arXiv:hep-ph/0305087 [hep-ph]].

\bibitem{Denner:2015yca}
A.~Denner and R.~Feger,
JHEP \textbf{11}, 209 (2015)
doi:10.1007/JHEP11(2015)209
[arXiv:1506.07448 [hep-ph]].

\bibitem{Kulesza:2015vda}
A.~Kulesza, L.~Motyka, T.~Stebel and V.~Theeuwes,
JHEP \textbf{03}, 065 (2016)
doi:10.1007/JHEP03(2016)065
[arXiv:1509.02780 [hep-ph]].

\bibitem{Broggio:2016lfj}
A.~Broggio, A.~Ferroglia, B.~D.~Pecjak and L.~L.~Yang,
JHEP \textbf{02}, 126 (2017)
doi:10.1007/JHEP02(2017)126
[arXiv:1611.00049 [hep-ph]].

\bibitem{Kulesza:2017ukk}
A.~Kulesza, L.~Motyka, T.~Stebel and V.~Theeuwes,
Phys. Rev. D \textbf{97} (2018) no.11, 114007
doi:10.1103/PhysRevD.97.114007
[arXiv:1704.03363 [hep-ph]].

\bibitem{Kulesza:2020nfh}
A.~Kulesza, L.~Motyka, D.~Schwartl\"ander, T.~Stebel and V.~Theeuwes,
Eur. Phys. J. C \textbf{80} (2020) no.5, 428
doi:10.1140/epjc/s10052-020-7987-6
[arXiv:2001.03031 [hep-ph]].

\bibitem{Broggio:2015lya}
A.~Broggio, A.~Ferroglia, B.~D.~Pecjak, A.~Signer and L.~L.~Yang,
JHEP \textbf{03}, 124 (2016)
doi:10.1007/JHEP03(2016)124
[arXiv:1510.01914 [hep-ph]].

\bibitem{vanDeurzen:2013xla}
H.~van Deurzen, G.~Luisoni, P.~Mastrolia, E.~Mirabella, G.~Ossola and T.~Peraro,
Phys. Rev. Lett. \textbf{111}, no.17, 171801 (2013)
doi:10.1103/PhysRevLett.111.171801
[arXiv:1307.8437 [hep-ph]].

\bibitem{Yu:2014cka}
Y.~Zhang, W.~G.~Ma, R.~Y.~Zhang, C.~Chen and L.~Guo,
Phys. Lett. B \textbf{738}, 1-5 (2014)
doi:10.1016/j.physletb.2014.09.022
[arXiv:1407.1110 [hep-ph]].

\bibitem{Frixione:2014qaa}
S.~Frixione, V.~Hirschi, D.~Pagani, H.~S.~Shao and M.~Zaro,
JHEP \textbf{09}, 065 (2014)
doi:10.1007/JHEP09(2014)065
[arXiv:1407.0823 [hep-ph]].

\bibitem{Frixione:2015zaa}
S.~Frixione, V.~Hirschi, D.~Pagani, H.~S.~Shao and M.~Zaro,
JHEP \textbf{06}, 184 (2015)
doi:10.1007/JHEP06(2015)184
[arXiv:1504.03446 [hep-ph]].

\bibitem{Hartanto:2015uka}
H.~B.~Hartanto, B.~Jager, L.~Reina and D.~Wackeroth,
Phys. Rev. D \textbf{91}, no.9, 094003 (2015)
doi:10.1103/PhysRevD.91.094003
[arXiv:1501.04498 [hep-ph]].

\bibitem{Mele:1990cw}
B.~Mele and P.~Nason,
Nucl. Phys. B \textbf{361}, 626-644 (1991)
[erratum: Nucl. Phys. B \textbf{921}, 841-842 (2017)]

\bibitem{Dawson:1997im}
S.~Dawson and L.~Reina,
Phys. Rev. D \textbf{57} (1998), 5851-5859
doi:10.1103/PhysRevD.57.5851
[arXiv:hep-ph/9712400 [hep-ph]].

\bibitem{Braaten:2015ppa}
E.~Braaten and H.~Zhang,
Phys. Rev. D \textbf{93} (2016) no.5, 053014
doi:10.1103/PhysRevD.93.053014
[arXiv:1510.01686 [hep-ph]].

\bibitem{Berman:1971xz}
S.~M.~Berman, J.~D.~Bjorken and J.~B.~Kogut,
Phys. Rev. D \textbf{4}, 3388 (1971)

\bibitem{Altarelli:1977zs}
G.~Altarelli and G.~Parisi,
Nucl. Phys. B \textbf{126}, 298-318 (1977)

\bibitem{Dokshitzer:1977sg}
Y.~L.~Dokshitzer,
Sov. Phys. JETP \textbf{46}, 641-653 (1977)

\bibitem{Gribov:1972ri}
V.~N.~Gribov and L.~N.~Lipatov,
Sov. J. Nucl. Phys. \textbf{15}, 438-450 (1972)
IPTI-381-71.

\bibitem{Melnikov:2004bm}
K.~Melnikov and A.~Mitov,
Phys. Rev. D \textbf{70}, 034027 (2004)
[arXiv:hep-ph/0404143 [hep-ph]].

\bibitem{Mitov:2004du}
A.~Mitov,
Phys. Rev. D \textbf{71}, 054021 (2005)
[arXiv:hep-ph/0410205 [hep-ph]].

\bibitem{Collins:1981uw}
J.~C.~Collins and D.~E.~Soper,
Nucl. Phys. B \textbf{194}, 445-492 (1982)
doi:10.1016/0550-3213(82)90021-9

\bibitem{Czakon:2021ohs}
M.~L.~Czakon, T.~Generet, A.~Mitov and R.~Poncelet,
[arXiv:2102.08267 [hep-ph]].

\bibitem{Mathematica}
Wolfram Research, Inc., \textit{Mathematica, Version 12.2}, Champaign, IL (2020).

\bibitem{FeynCalc}
V. Shtabovenko, R. Mertig and F. Orellana, Comput. Phys. Commun. 256 (2020) 107478, arXiv:2001.04407; V. Shtabovenko, R. Mertig and F. Orellana, Comput. Phys. Commun. 207 (2016) 432-444, arXiv:1601.01167; R. Mertig, M. Böhm, and A. Denner, Comput. Phys. Commun. 64 (1991) 345-359.

\bibitem{Anastasiou:2002yz}
C.~Anastasiou and K.~Melnikov,
Nucl. Phys. B \textbf{646}, 220-256 (2002)
doi:10.1016/S0550-3213(02)00837-4
[arXiv:hep-ph/0207004 [hep-ph]].

\bibitem{Anastasiou:2002wq}
C.~Anastasiou and K.~Melnikov,
Phys. Rev. D \textbf{67}, 037501 (2003)
doi:10.1103/PhysRevD.67.037501
[arXiv:hep-ph/0208115 [hep-ph]].

\bibitem{Anastasiou:2002qz}
C.~Anastasiou, L.~J.~Dixon and K.~Melnikov,
Nucl. Phys. B Proc. Suppl. \textbf{116}, 193-197 (2003)
doi:10.1016/S0920-5632(03)80168-8
[arXiv:hep-ph/0211141 [hep-ph]].

\bibitem{Anastasiou:2003yy}
C.~Anastasiou, L.~J.~Dixon, K.~Melnikov and F.~Petriello,
Phys. Rev. Lett. \textbf{91}, 182002 (2003)
doi:10.1103/PhysRevLett.91.182002
[arXiv:hep-ph/0306192 [hep-ph]].

\bibitem{Anastasiou:2003ds}
C.~Anastasiou, L.~J.~Dixon, K.~Melnikov and F.~Petriello,
Phys. Rev. D \textbf{69}, 094008 (2004)
doi:10.1103/PhysRevD.69.094008
[arXiv:hep-ph/0312266 [hep-ph]].

\bibitem{Tkachov:1981wb}
F.~V.~Tkachov,
Phys. Lett. B \textbf{100}, 65-68 (1981)
doi:10.1016/0370-2693(81)90288-4

\bibitem{Chetyrkin:1981qh}
K.~G.~Chetyrkin and F.~V.~Tkachov,
Nucl. Phys. B \textbf{192}, 159-204 (1981)
doi:10.1016/0550-3213(81)90199-1

\bibitem{Smirnov:2019qkx}
A.~V.~Smirnov and F.~S.~Chuharev,
Comput. Phys. Commun. \textbf{247 }, 106877 (2020)
doi:10.1016/j.cpc.2019.106877
[arXiv:1901.07808 [hep-ph]].

\bibitem{Huber:2005yg}
T.~Huber and D.~Maitre,
Comput. Phys. Commun. \textbf{175}, 122-144 (2006)
doi:10.1016/j.cpc.2006.01.007
[arXiv:hep-ph/0507094 [hep-ph]].

\bibitem{Huber:2007dx}
T.~Huber and D.~Maitre,
Comput. Phys. Commun. \textbf{178}, 755-776 (2008)
doi:10.1016/j.cpc.2007.12.008
[arXiv:0708.2443 [hep-ph]].

\bibitem{Maitre:2005uu}
D.~Maitre,
Comput. Phys. Commun. \textbf{174}, 222-240 (2006)
doi:10.1016/j.cpc.2005.10.008
[arXiv:hep-ph/0507152 [hep-ph]].

\bibitem{Maitre:2007kp}
D.~Maitre,
Comput. Phys. Commun. \textbf{183}, 846 (2012)
doi:10.1016/j.cpc.2011.11.015
[arXiv:hep-ph/0703052 [hep-ph]].

\bibitem{Kotikov:1990kg}
A.~V.~Kotikov,
Phys. Lett. B \textbf{254}, 158-164 (1991)
doi:10.1016/0370-2693(91)90413-K

\bibitem{Kotikov:1991hm}
A.~V.~Kotikov,
Phys. Lett. B \textbf{259}, 314-322 (1991)
doi:10.1016/0370-2693(91)90834-D

\bibitem{Bern:1992em}
Z.~Bern, L.~J.~Dixon and D.~A.~Kosower,
Phys. Lett. B \textbf{302}, 299-308 (1993)
[erratum: Phys. Lett. B \textbf{318}, 649 (1993)]
doi:10.1016/0370-2693(93)90400-C
[arXiv:hep-ph/9212308 [hep-ph]].

\bibitem{Bern:1993kr}
Z.~Bern, L.~J.~Dixon and D.~A.~Kosower,
Nucl. Phys. B \textbf{412}, 751-816 (1994)
doi:10.1016/0550-3213(94)90398-0
[arXiv:hep-ph/9306240 [hep-ph]].

\bibitem{Kotikov:1993zf}
A.~V.~Kotikov,
JINR-E2-93-414.

\bibitem{Fleischer:1997bw}
J.~Fleischer, A.~V.~Kotikov and O.~L.~Veretin,
Phys. Lett. B \textbf{417}, 163-172 (1998)
doi:10.1016/S0370-2693(97)01195-7
[arXiv:hep-ph/9707492 [hep-ph]].

\bibitem{Gehrmann:1999as}
T.~Gehrmann and E.~Remiddi,
Nucl. Phys. B \textbf{580}, 485-518 (2000)
doi:10.1016/S0550-3213(00)00223-6
[arXiv:hep-ph/9912329 [hep-ph]].

\bibitem{Henn:2013pwa}
J.~M.~Henn,
Phys. Rev. Lett. \textbf{110}, 251601 (2013)
doi:10.1103/PhysRevLett.110.251601
[arXiv:1304.1806 [hep-th]].

\bibitem{Duhr:2019tlz}
C.~Duhr and F.~Dulat,
JHEP \textbf{08}, 135 (2019)
doi:10.1007/JHEP08(2019)135
[arXiv:1904.07279 [hep-th]].

\bibitem{Goncharov.A.B.:2009tja}
A.~B.~Goncharov,
[arXiv:0908.2238 [math.AG]].

\bibitem{Goncharov:2010jf}
A.~B.~Goncharov, M.~Spradlin, C.~Vergu and A.~Volovich,
Phys. Rev. Lett. \textbf{105}, 151605 (2010)
doi:10.1103/PhysRevLett.105.151605
[arXiv:1006.5703 [hep-th]].

\bibitem{Duhr:2011zq}
C.~Duhr, H.~Gangl and J.~R.~Rhodes,
JHEP \textbf{10}, 075 (2012)
doi:10.1007/JHEP10(2012)075
[arXiv:1110.0458 [math-ph]].

\bibitem{DiVita:2014pza}
S.~Di Vita, P.~Mastrolia, U.~Schubert and V.~Yundin,
JHEP \textbf{09}, 148 (2014)
doi:10.1007/JHEP09(2014)148
[arXiv:1408.3107 [hep-ph]].

\bibitem{deFlorian:2016spz}
D.~de Florian \textit{et al.} [LHC Higgs Cross Section Working Group],
doi:10.23731/CYRM-2017-002
[arXiv:1610.07922 [hep-ph]].

\bibitem{JaxoDraw}
The Feyman diagrams were made using JaxoDraw: D.~Binosi, J.~Collins, C.~Kaufhold and L. Theussl,
Comput. Phys. Commun. 180, 1709-1715 (2009),
[arXiv:0811.4113].

\end{thebibliography}
\end{document}